\documentclass[aps,a4paper,10pt,twocolumn,nofootinbib,eqsecnum,showpacs]{revtex4} 
\usepackage[T1]{fontenc}
\usepackage{mathptmx}
\usepackage{datetime}
\usepackage{amsmath}
\usepackage{amsfonts}
\usepackage{mathrsfs}
\usepackage[mathscr]{euscript}
\usepackage{mathtools}
\usepackage{textcomp} 
\usepackage{fancyhdr}
\usepackage{colordvi}
\usepackage{epsfig}
\usepackage{color}
\usepackage{bm}
\usepackage{harpoon}
\usepackage{stmaryrd}
\usepackage{cancel}
\usepackage{tensor}
\usepackage[normalem]{ulem} 
\def\overstrike#1#2{{\setbox0\hbox{$#2$}\hbox to \wd0{\hss
    $#1$\hss}\kern-\wd0\box0}}

        \DeclareMathOperator{\grad}{\nabla}

\newcommand{\dualop}{\star}
\newcommand{\dualF}{{\dualop\!\! F}} 
\newcommand{\dualchi}{{\dualop\!\! \chi}} 
\renewcommand{\Vec}[1]{\textbf{#1}}

\newcommand{\Dffsn}{\mathfrak{D}}
\newcommand{\CSQchar}{C}
\newcommand{\CSQ}{\mathfrak{\CSQchar}}

\newdateformat{yymmdddate}{\THEYEAR/\twodigit{\THEMONTH}/\twodigit{\THEDAY}}

\newcommand{\XDOI}[1]{\href{http://dx.doi.org/#1}{doi:#1}}
\newcommand{\XARXIV}[1]{\href{http://arxiv.org/abs/#1}{arXiv:#1}}
\newcommand{\XWEB}[1]{\href{#1}{#1}}

\newcommand{\shadow}[1]{{{#1}_s}}

\newcommand{\manifold}[1]{\mathcal{#1}}

\newcommand{\indevice}[1]{\tilde{#1}}
\newcommand{\indesign}[1]{\hat{#1}}
\newcommand{\xinverse}[1]{\bar{#1}}

\newcommand{\TEXTDEL}[1]{}

\newcommand{\TEXTNEW}[2]{#1}

\def\mReals{\mathbb{R}}    
\def\mCircles{\mathbb{S}}  

\newcommand{\pseudosection}[1]{~\\
\textbf{#1}}

\def\hhzfield{\Psi}

\def\Morph{\varphi}
\def\MorphI{(\Morph^{-1})}

\def\Point{\mathcal{P}}
\def\coder{{\boldsymbol \nabla}}

\def\newGamma{{\indevice{\Gamma}}}
\def\pstar{\ast}

\def\gMetric{g}
\def\gCoMetric{\mathfrak{\gMetric}}

\def\manM{{\manifold{M}}}
\def\subM{{\manifold{U}}}
\def\manDes{{\indesign{\manM}}}
\def\subDes{{\indesign{\subM}}}
\def\manDev{{\indevice{\manM}}}

\def\figwidthFull{0.80\columnwidth}
\def\figwidthHalf{0.44\columnwidth}

\begin{document}

\title{Generalized Transformation Design: metrics, speeds, and diffusion}

\author{Paul Kinsler$^{1,2,3}$}
\homepage[]{https://orcid.org/0000-0001-5744-8146}
\email[\hphantom{.}~]{Dr.Paul.Kinsler@physics.org}

\author{Martin W. McCall$^3$}
\homepage[]{https://orcid.org/0000-0003-0643-7169}

\affiliation{$^1$Cockcroft Institute, Keckwick Lane,
  Daresbury,
  WA4 4AD,
  United Kingdom.}

\affiliation{$^2$Physics Department,
  Lancaster University,
  Lancaster LA1 4YB,
  United Kingdom.}

\affiliation{$^3$
  Blackett Laboratory, Imperial College London,
  Prince Consort Road,
  London SW7 2AZ,
  United Kingdom.}

\begin{abstract}

We show that a unified and maximally generalized
 approach to spatial transformation design
 is possible,
 one that encompasses all second order waves, 
 rays,
 and diffusion processes in anisotropic media. 
Until the final step,
 it is unnecessary to specify the physical process
 for which a specific transformation design is to be implemented. 
The principal approximation is the neglect of wave impedance, 
 an attribute that plays no role in ray propagation,
 and is therefore irrelevant for pure ray devices;
 another constraint is that for waves
 the spatial variation in material parameters
 needs to be sufficiently small compared with the wavelength. 
The key link between our general formulation
 and a specific implementation is how the spatial metric
 relates to the speed of disturbance in a given medium,
 whether it is electromagnetic, acoustic, or diffusive.
Notably,
 we show that our generalised ray theory,
 in allowing for anisotropic indexes (speeds),
 generates the same predictions as does a wave theory,
 and the results are closely related to those for diffusion processes.

\end{abstract}





\lhead{\includegraphics[height=5mm,angle=0]{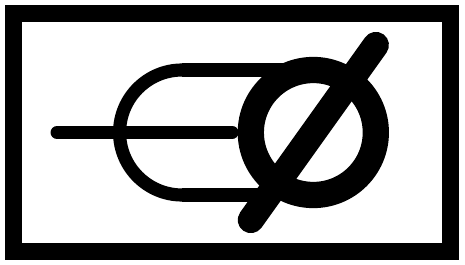}~~RAYTAIL}
\chead{Generalized Transformation Design}
\rhead{
\href{mailto:Dr.Paul.Kinsler@physics.org}{Dr.Paul.Kinsler@physics.org}\\
\href{http://www.kinsler.org/physics/}{http://www.kinsler.org/physics/}
}

\date{\today}
\maketitle
\thispagestyle{fancy}


%
\section{Introduction}
\label{S-Introduction}

Transformation design (T-Design) is a way of constructing devices
 based directly on a mathematical specification.
The essence of the idea is that it lets us shift waves,
 rays, 
 or other excitations around inside the device, 
 while altering the way they propagate, 
 so that the outside world sees no changes.
Here we make this more mathematically precise through a two stage process: 
 first by defining a "morphism" picture that applies equally to all cases, 
 and then a second step that matches the morphism picture
 to the specific physical system.

The most notable example of T-Design is that 
 of electromagnetic cloaking,
 which has now 
 been with us for almost ten years
 \cite{Pendry-SS-2006sci,Leonhardt-2006sci}.
It has been recently revitalized by the introduction
 of the concept of space-time cloaking
 \cite{McCall-FKB-2011jo,Kinsler-M-2014adp-scast,Gratus-KMT-2016njp-stdisp}
 and its variety of implementations
 \cite{Fridman-FOG-2012nat,Lukens-MLW-2014o,Chremmos-2014ol,Jabar-BA-2015lp}
In light of the many variants of spatial cloaking, 
 and of applications in acoustics
 \cite{Zhang-XF-2011prl,Sanchis-GLCMCS-2013prl}
  and diffusion of heat or light
 \cite{Guenneau-AV-2012oe,Schittny-NKBNW-2015o}, 
 it is worth considering how to combine 
 these different applications and approaches
 into a unified T-Design scheme, 
 at least to the extent possible.
Some progress has been made in that regard, 
 but with a firm focus on wave mechanics expressed in a first-order form
 \cite{Kinsler-M-2014pra}.

Here we take an exclusively second order approach
 in which a subset of variables
 satisfying a system of first order equations
 are expressed as a single second order equation.
For example,
 instead of examining transformations of Maxwell's equations
 in the field vectors $\Vec{E}, \Vec{B}, \Vec{D}$ and $\Vec{H}$, 
 we will consider just the simplest (Helmholtz-like) 
 second order wave equation
 a single field component, 
 e.g. just $E$, or just $B$.
Although less detailed than first-order approaches, 
 notably in the way impedance is ignored,
 the second-order approach has some significant advantages.
In any case, 
 when it comes to 
 actual 
 \emph{Transformation devices} (T-devices) --
 often impedances are ruthlessly ignored or rescaled
 to suit the technological demands
 of the programme.
In that sense including impedance could be considered
 somewhat over careful.
T-devices in which no attempt is made to control scattering
 or reflections from impedance mismatches
 have a performance left hostage 
 to the physics of the wave or ray 
 they attempt to manipulate.
Notably, 
 we expect that in principle electromagnetic (EM)
 T-devices are less imperfect
 than acoustic ones
 \cite{Kinsler-M-2014pra}.

We start by taking a
 second order wave equation
 as describing a given system,
 and see how it can be modified to allow for anisotropy.
Since, 
 largely,
 any wave type can be modelled this way,
 this is not a particularly stringent restriction.
We then show how the equation can be recast into a covariant form, 
 where the covariant derivative is that associated 
 with the space's underlying metric, $g_{\alpha\beta}$.
Transformation design
 is then described as a mathematical morphism
 between a reference or ``design'' solution and a T-device application; 
 the extraction of material parameters from the morphed/transformed metric --
 for whatever physical system is of interest --
 is then solely a problem of calculation. 

In section \ref{S-theory} we present the basic wave,
 ray, 
 and diffusion machinery
 in the context of our second-order approach,
 and how, 
 for a given choice of physical system, 
 the material properties map onto the effective metric.
In section \ref{S-transformations} we show how the metric morphs 
 under transformation, 
 and in section \ref{S-examples} we give examples.
Finally, 
 in section  \ref{S-conclusions} we summarize our results.

%
\section{Waves, Rays, and Diffusions}
\label{S-theory}

In what follows we will generalise 
 the common second order wave equation approach for T-Design
 to allow for anisotropy of the propagation in the simplest possible way.
We then demonstrate how this generalization
 allows a unified process for designing T-devices
 for almost any sort of wave, 
 ray, 
 or diffusion.
This is because all these types of processes
 can have their mathematical expression and behaviour mapped on to the 
 metric seen by the process, 
 so that a {transformation} of a metric is sufficent
 to determine the necessary material parameters
 for the chosen T-device.

%
\subsection{Waves}
\label{S-theory-waves}

The most general type of second order wave model
 is given by the covariant wave equation 
 on a manifold where the spatial part of the metric ${\gMetric}\indices{_{ij}}$
 has its inverse counterpart ${\gMetric}\indices{^{ij}}$.
In indexed notation using the Einstein summation convention, 
 with Greek indices spanning $\{ t,x,y,x \}$ and
 Latin ones spanning $\{x,y,z\}$,
 and treating the time coordinate $t$ 
 separately from space,
 this is 
~
\begin{align}
  \hhzfield\indices{^{;\mu}_{;\mu}}
=
  \grad_i 
   \left(
     {\gMetric}\indices{^{ij}}
      \grad_j
      \hhzfield
   \right)
 +
  {\gMetric}\indices{^{tt}}
  \partial_t^2 
   \hhzfield
&=
0
,
\label{eqn-covar-2owe}
\end{align}
where $\gMetric\indices{^{\mu\nu}}$
 is inverse to $\gMetric\indices{_{\mu\nu}}$.
The separation between space and time
 is justified when we are only interested in spatial morphisms.
Here both ``$;\mu$'' and $\grad_\mu$ denote covariant derivatives, 
 whereas $\partial_\mu$ are partial derivatives.
This equation is for a scalar field $\hhzfield$, 
 but the generalization to other types of waves is straightforward.
Note that for wave processes, 
 we have ${\gMetric}_{tt}={\gMetric}^{tt}=-1$.

Equation \eqref{eqn-covar-2owe} can be compared
 to the standard wave equation
 for a field in a homogeneous isotropic medium
 (e.g. consider light travelling in an ordinary block of glass).
~
\begin{align}
  \left[
    \sum_i
    \partial_i
    \left(
      v^2
      \partial_i
    \right)
   -
    \partial_t^2
  \right]
  \hhzfield(\Vec{r},t)
&=
  0
.
\label{eqn-basic-nabla2Ec2}
\end{align}
This non-covariant form of the second order wave equation
 is posed in Euclidean coordinates
 in which there is no need to distinguish
 between co-variant and contra-variant indices. 
Placing the homogeneous $v^2$ wave property between the
 two spatial derivatives facilitates
 comparison with the covariant form of \eqref{eqn-covar-2owe}.
The use of a squared property is also worthy of note -- 
 it has also been argued that for electromagnetic waves in media, 
 the 
 refractive index squared is a much more relevant quantity 
 than index $n$ in optical propagation \cite{Kinsler-2009pra}, 
 in particular as to how it affects the best definition of wavevector
 in the presence of significant gain or loss.

Comparing \eqref{eqn-covar-2owe}
 and \eqref{eqn-basic-nabla2Ec2}
 we see that they have a very similar form --
 the difference being that the isotropic and scalar speed squared ($v^2$)
 has been replaced by
 the potentially anisotropic inverse spatial metric ${\gMetric}\indices{^{ij}}$.
Thus
 the somewhat abstract notion of the inverse of the metric on a manifold 
 can be replaced by the concrete and intuitive notion of
 a speed (squared) matrix
 (which we will denote
  using the Fraktur `\CSQchar' character as
 ${\CSQ}^{ij}$).

Allowing anisotropy
 by considering materials with a speed matrix ${\CSQ}^{ij}$
 rather than an isotropic $v^2$
 is crucial for the field of Transformation Media: 
 the process of designing T-devices
 relies on the introduction of material properties
 that are both anisotropic and inhomogeneous
 \cite{Kinsler-M-2015pnfa-tofu}.
Even the simplest possible transformation -- 
 a single axis compression -- 
 induces anisotropy, 
 and therefore
 any general transformation theory 
 must incorporate it.
The sole exception is for T-devices designed by means of conformal maps
 \cite{Leonhardt-2006sci,Sarbort-T-2012jo},
 that produce,
 for example,
 cloaks 
 that work in just two dimensions for a single specific orientation, 
 and are therefore of limited utility. 
The covariant wave equation can now be expressed in terms of ${\CSQ}^{ij}$
 as
~
\begin{align}
  \left[
    \grad_i
    {\CSQ}\indices{^{ij}}
    \grad_j
   -
    \partial_t^2
  \right]
  \hhzfield(\Vec{r},t)
&=
  0
.
\label{eqn-basic-nabla2Ec2aniso}
\end{align}

To summarize:
 we assume that \emph{any} wave-like excitation, 
 appropriately specified, 
 can be described by the second order wave equation.
On this basis we can regard, 
 in well-founded, 
 but somewhat restricted terms, 
 the second order wave equation as the defining
 description of wave processes -- 
 i.e. any wavelike excitation of a field or material 
 follows (in some suitable limit)
 an archetypical Helmholtz-like formula
 with material properties contained in a speed-squared matrix ${\CSQ}^{ij}$.

%
\subsection{Rays}
\label{S-theory-ray}

In the short-wavelength (eikonal) limit
 the second order wave equation reduces to a ray equation. 
Rays,
 in which all sense of wave amplitude or polarization are lost, 
 and only the direction of propagation retained, 
 are \emph{geodesics} with respect to the space in which they travel. 
Optical rays,
 for example,
 traversing an inhomogeneous isotropic medium,
 extremize the optical path length (OPL) according to Fermat's principle
~
\begin{align}
  \delta (OPL) 
&=
  \delta 
  \int_A^B 
  \left(
    {\gMetric}\indices{_{ij}}
    \frac{dx^i}{ds}\frac{dx^j}{ds}
  \right)^{1/2}
  ds
~~=~~ 0
\label{eqn-geodesic}
,
\end{align}
where the optical metric ${\gMetric}\indices{_{ij}}$
 is given in Cartesians by $g_{ij}=n^2\delta_{ij}$. 
The resulting geodesic equation\footnote{An 
 alternative expression in terms of 
 two coupled first-order pieces
 is
~
$
  \frac{d v^i}{d\lambda}
=
 -
  \Gamma\indices{^{i}_{jk}}
  v^j 
  v^k
,
\quad
  \frac{d x^i}{d\lambda}
=
  v^i
.$
}
 is
~
\begin{align}
  \frac{d^2 x^i}{d\lambda^2}
 +
  \Gamma\indices{^{i}_{jk}}
  \frac{dx^j}{d\lambda}
  \frac{dx^k}{d\lambda}
&=
  0
\label{eqn-geodesic-std}
,
\end{align}
 where the connection coefficients in Cartesians are given by
~
\begin{align}
  \Gamma\indices{^{i}_{jk}}
&=
  \left(
    \delta\indices{^{i}_{j}}\partial_k
   +
    \delta\indices{^{i}_{k}}\partial_j
   -
    \delta^{im}\delta_{jk} \partial_m
  \right)
  \left[\ln (n)\right]
&=
  0
\label{eqn-optical-connection}
.
\end{align}
It is straightforward to show that
 \eqref{eqn-geodesic-std} and \eqref{eqn-optical-connection}
 are equivalent to
 the standard ray equation\footnote{Using
   the fact that $(d{\bf r}/ds)\cdot(d{\bf r}/ds)=1$,
   \eqref{eqn-ray-std} can be
   straightforwardly manipulated to
   $\frac{d^2{\bf r}}{ds^2} +
     +\left(\frac{d{\bf r}}{ds}\cdot\nabla \ln n\right)\frac{d{\bf r}}{ds} 
     -\nabla \ln n \left(\frac{d{\bf r}}{ds}\right)
     \cdot
     \left(\frac{d{\bf r}}{ds}\right)
    = 0$,
   from which Eqs. \eqref{eqn-geodesic-std}
   and \eqref{eqn-optical-connection} follow.}
~
\begin{align}
  \frac{d}{ds}
  \left(
    n
    \frac{d{x^i}}{ds}
  \right)
&=
  \partial_i n
\label{eqn-ray-std}
.
\end{align}

A uniform medium,
 characterised by a homogeneous index $n$ yields the `straight lines' of Cartesian space,
 $x^i=x^i_0+nv^i_0s$.
By identifying rays as geodesics
 with respect an arbitrary spatial metric $ds^2=g_{ij}dx^idx^j$,
 the ray limit of the covariant wave equation \eqref{eqn-covar-2owe}
 is again the geodesic equation, 
 \eqref{eqn-geodesic},
 where now the connection coefficients are just given by the standard formula
~
\begin{align}
  \Gamma\indices{^{i}_{jk}}
&=
  \frac{1}{2}
  {\gMetric}\indices{^{im}}
  \left[
    \partial_k {\gMetric}\indices{_{jm}}
   +
    \partial_j {\gMetric}\indices{_{km}}
   -
    \partial_m {\gMetric}\indices{_{jk}}
  \right]
\label{eqn-connection-Gammas}
.
\end{align}
In fact, by making the usual
 short-wavelength and ray-limit approximations
 to \eqref{eqn-covar-2owe} the following
 covariant ray equation is obtained
 as the genereralization of \eqref{eqn-ray-std}:
~
\begin{align}
  \frac{d}{ds}
  \left(
    {\gMetric}_{ij} \frac{dx^j}{ds}
  \right)
&=
  \frac{1}{2}
  {\gMetric}_{mn,i}
  \frac{dx^m}{ds}\frac{dx^n}
  {ds}
.
\label{eqn-genericeikonal}
\end{align}
This equation can be manipulated to yield 
 \eqref{eqn-geodesic-std} and \eqref{eqn-connection-Gammas}.
The crucial thing to note here is that just as for the 
 second order wave equation, 
 the controlling property for the geodesics --
 for motion or transport across the manifold -- 
 is the metric.

The key idea in what follows
 is that morphing geodesics from one space to another
 amounts to a mapping of the metric from the design space to the device space. In turn,
 once the medium parameters are related to the metric in the design space,
 they are determined in the device space. 
A typical progression is to take the scaled Cartesian metric
 ${\CSQ}^{ij}=v^2\delta^{ij}$ in the design space,
 and infer the required anisotropic medium parameters ${\CSQ}^{ij}$
 in the device space.

%
\subsection{Diffusions}

We will consider two types of equation under the heading 
 ``diffusion''.
Firstly, 
 although not usually regarded as a diffusion equation,
 we can flip the sign on the time derivative term
 of the second order wave equation, 
 i.e. set ${\gMetric}^{00}=+1$ in \eqref{eqn-covar-2owe}.
To emphasize that we intend to treat diffusion-like processes, 
 we replace the inverse metric ${\gMetric}^{ij}$ 
 not with a speed squared matrix ${\CSQ}^{ij}$ 
 but a diffusion matrix ${\Dffsn}^{ij}$, 
 so that 
~
\begin{align}
  \left[
    \partial_i
      {\Dffsn}\indices{^{ij}}
    \partial_j
   +
    \partial_t^2
  \right]
  \hhzfield(\Vec{r},t)
&=
  0
.
\label{eqn-basic-nabla2diffusions}
\end{align}
Thus  any deductions made on the basis of spatial transformations
 for \eqref{eqn-covar-2owe},
 apply also to \eqref{eqn-basic-nabla2diffusions}.

%

More traditional diffusion equations,
 or even Schr\"odinger type equations,
 which contain only a first order time derivative
 can also be treated under the same 
 machinery outlined above.
This allows us to bring calculations
 such as that of the heat diffusion cloak \cite{Guenneau-AV-2012oe}
 into the unified picture described in this work.

To demonstrate how the transformation schemes for 
 waves and diffusions follow the same process, 
 {and therefore fit into our generalized scheme,}
 we first define a ``shadow'' wave equation for a field $S$,
 intended to mimic a diffusion equation in some suitable limit\footnote{This
  shadow system, 
  if necessary, 
  can be considered to follow the same pair of first order equations
  as p-acoustics; 
  see sec. \ref{S-theory-examples}}.
For this we require an added $\alpha^2 S$ source term on the RHS, 
 and to define a new diffusive field quantity $\shadow{\hhzfield}$
 related to the shadow field $S$ with $S = \shadow{\hhzfield} \exp(-\zeta t)$.
This means that 
~
\begin{align}
  \left[
    \partial_i
      {\CSQ}\indices{^{ij}}
    \partial_j
   +
    \partial_t^2
  \right]
  {S}(\Vec{r},t)
&=
  \alpha^2 {S}(\Vec{r},t)
,
\\
  \left[
    \partial_i
      {\CSQ}\indices{^{ij}}
    \partial_j
   +
    \zeta^2
   - 
    2
    \zeta
    \partial_t
   +
    \partial_t^2
  \right]
  \shadow{\hhzfield}(\Vec{r},t)
&=
  \alpha^2 \shadow{\hhzfield}(\Vec{r},t)
.
\label{eqn-diffusion-nabla2}
\end{align}

We then assume that there there exists a value of $\zeta$
 sufficiently large that $\shadow{\hhzfield}$ will always vary slowly
 compared to $\exp(-\zeta t)$, 
 i.e. 
~
\begin{align}
  \left|\zeta\right| 
\gg
  \left| \frac{\partial_t \shadow{\hhzfield}}{\shadow{\hhzfield}} \right|
.
\label{eqn-diffusion-approx}
\end{align}
We can then drop the negligible $\partial_t^2 \shadow{\hhzfield}$ term
 from the above equation.
If we also match the source-like parameter $\alpha$
 to $\zeta$ with $\alpha^2 = \zeta^2$, 
 and setting ${\Dffsn}\indices{^{ij}}={\CSQ}\indices{^{ij}}/2\zeta$
 we obtain
~
\begin{align}
  \left[
    \partial_i
      {\Dffsn}\indices{^{ij}}
    \partial_j
   -
    \partial_t
  \right]
  \shadow{\hhzfield}(\Vec{r},t)
&=
  0
,
\label{eqn-diffusion-reorg2}
\end{align}
 which has the same form
 as a diffusion or Schr\"odinger equation\footnote{Note that 
   there are more systematic ways of converting between second order 
   and Schr\"odinger equation forms \cite{Kinsler-2013arxiv-kg2schro}.}.
Treated as a Schr\"odinger equation,
 \eqref{eqn-diffusion-reorg2} incorporates
 anisotropic effective mass
 appropriate for particles in anisotropic periodic potential
 as found in crystals \cite{Kittel-ISSP}.

As a result of the above calculation, 
 given any diffusion/ Schr\"odinger equation
 of the form \eqref{eqn-diffusion-reorg2}
 we can choose a sufficiently large $\zeta$, 
 calculate the effective ${\CSQ}\indices{^{ij}} = 2 \zeta {\Dffsn}\indices{^{ij}}$, 
 transform it in the way described below 
 to get the desired device's
 effective speed squared $\indevice{\CSQ}\indices{^{ij}}$, 
 and then the desired device's diffusion
 $\indevice{\Dffsn}\indices{^{ij}} = \indevice{\CSQ}\indices{^{ij}}/2\zeta$.

Note that the properties of field $\hhzfield$ and parameter $\zeta$
 are only constrained by the need to satisfy the approximation 
 of \eqref{eqn-diffusion-approx}; 
 they are merely used to define a ``shadow'' wave equation
 which is not intended to have a direct physical interpretation.
Crucially,
 since $\zeta$ is a simple scalar
 and ${\Dffsn}\indices{^{ij}}$ is directly proportional
 to ${\CSQ}\indices{^{ij}}$,
 we can just transform ${\Dffsn}\indices{^{ij}}$ directly
 to determine the necessary T-device diffusion (or potential) properties.

%
\subsection{Making the metric}
\label{S-theory-examples}

Although general equations 
 such as \eqref{eqn-covar-2owe}, 
 \eqref{eqn-geodesic-std},
 or \eqref{eqn-diffusion-reorg2}
 are invaluable starting points,
 in general we need to justify their use
 based on particular physical models.
These models then will show us how constitutive or material parameters
 will combine to form the effective metric
 for that type of wave
 -- at least in the limit where the behaviour
 is straightforward enough to be safely characterized in such a way.
Here we will do this following 
 our previous work which attempted a unification 
 of T-optics and T-acoustics \cite{Kinsler-M-2014pra}; 
 i.e. we derive second order wave equations 
 directly from a generalization 
 of a p-acoustic model \cite{Kinsler-M-2014pra}, 
 as well as from electromagnetism (EM).
We use p-acoustics in place
 of some more specific acoustic model, 
 most notably because in the limits under consideration here, 
 many acoustical models reduce to a second order form 
 that can be easily represented within the p-acoustic framework. 
Further,
 the formulation of p-acoustics makes it an ideal vehicle 
 for incorporation of compatible acoustic systems
 within our generalized transformation design scheme.
However, 
 note that when using simplified models such as p-acoustics
 to represent mechanical systems under transformation,
 some caution remains necessary (see e.g. \cite{Norris-2012apl}).

{One critical point about any derivation that proceeds from the 
 original first order equations for 
 the models given below
 is that we assume the underlying constitutive parameters
 have a limited spatial and temporal dependence
 and vary slowly with respect to the wavelength; 
 this constraint is in accordance with that 
 specifying that the determinant of the metric undergoes negligible change
 in the simplifed covariant wave equation.}

\pseudosection{p-Acoustics:~}
In the case of generalized p-acoustics, 
 the equations for 
 velocity field ${v}^i$,
 and momentum density ${V}^i$, 
 in combination with amplitude $P$ and stress ${p}^{ij}$
 can be written in an indexed form.
In the rest frame of the acoustic medium,
 we have
~
\begin{align}
  \partial_t P       &= -\partial_i {v}^i, &~~
  \partial_t {V}^m   &= -\partial_n {p}^{mn}, \\
  {p}^{jk} &= -{\kappa}^{jk} P &~~
  {V}^m    &=  {\rho}\indices{^{m}_{i}} {v}^i, 
 \end{align}
 where we also need to know that 
 {there exist inverses
  $\xinverse{\kappa}_{rs}$ and $\xinverse{\rho}\indices{_{m}^{i'}}$
  such that}
 $ {\kappa}^{rs} \xinverse{\kappa}_{rs} = 1$,
 and
 $ \xinverse{\rho}\indices{_{m}^{i'}} {\rho}\indices{^{m}_{i}}
    = \delta\indices{^{i'}_{i}}$. 
As expected,  
 the momentum density field is related to the velocity field
 by a matrix of mass-density parameters ${\rho}\indices{^{m}_{i}}$.

For ordinary p-acoustics,
 $P$ is a scalar field representing the local population, 
 and ${\kappa}^{ij} = \kappa_o \delta^{ij}$ repesents the bulk modulus; 
 as a result ${p}^{kl} = p_o \delta^{kl}$
 so that $p_o$ is a pressure field.
There is also a version of p-acoustics
 that mimics pentamode materials \cite{Norris-2008rspa},
 where the modulus ${\kappa}^{ij}$ is a symmetric matrix
 but ${\rho}^{ij} = \rho_o \delta^{ij}$.
Most generally,
 p-acoustics allows the case where
 ${\kappa}^{ij}$ and ${p}^{jk}$ can be
 (at least in principle) any symmetric matrix; 
 in this case $P$ represents the amplitude of 
 an oscillating stress field
 whose orientation is determined by ${\kappa}^{jk}$, 
 and where the restoring stress is ${p}^{jk}$ is proportional to $P$.

The usual process for generating a second order wave equation
 then leads straightforwardly to 
~
\begin{align}
  \partial_t^2 P
&=
  \partial_i {\CSQ}^{ij} \partial_j P
\end{align}
 with a speed-squared matrix ${\CSQ}^{is}$ that
 depends on the bulk modulus $\kappa$
 and the mass density
 i.e.
~ 
\begin{align}
  {\CSQ}^{ij}
&=
  \xinverse{\rho}\indices{_n^i} \kappa^{nj}
,
\end{align}
 where $\xinverse{\rho}\indices{_n^i} {\rho}\indices{^m_i} = \delta\indices{^m_n}$.

\pseudosection{Electromagnetism:~} 
One approach to wave electromagnetics would be simply to write down
 a refractive index matrix,
 and use this as the basis for a spatial metric. 
However,
 in transformation optics this strictly applies only when transformations
 of the dielectric tensor $\boldsymbol \epsilon$ are matched
 to transformations of the permeability tensor $\boldsymbol \mu$.  
Here we take a more basic approach and derive a speed-squared matrix
 from Maxwell equations in tensor form. 
The presentation emphasises the structural similarity
 between p-acoustics and electromagnetism.

We can rewrite the vector Maxwell equations in an indexed form
 which incorporates the vector cross product
 by turning the electric and magnetic fields $E^i$ and $H^l$
 into antisymmetrized matrices $e^{kj}$ and $-h^{mn}$; 
 these also match up with the purely spatial parts of the 
 EM tensors\footnote{Here, 
 we use the Hodge dual operator $\dualop$ 
 (see e.g. \cite{Flanders1963})
 to convert the usual EM $F$ tensor
 into a more vector-notation friendly $\dualF$.}
 $\dualF$ and ${G}$.
The indexed equations, 
 which also can be extracted from a matrix representation 
 of the covariant tensor Maxwell equations
 (see e.g. \cite{Kinsler-M-2014pra}),
 can be written, 
~
\begin{align}
  \partial_t {B}^j &= -\partial_k {e}^{kj}, &~~
  \partial_t {D}^m      &= -\partial_n {h}^{mn} \\
  {h}^{mn}  &= {\eta}\indices{^{mn}_{j}} {B}^j, &~~
  {D}^m     &= {\epsilon}\indices{^{m}_{kl}} {e}^{kl},
 \end{align}
 where 
 $\xinverse{\eta}\indices{_{mn}^{j'}} {\eta}\indices{^{mn}_{j}} = \delta\indices{^{j'}_{j}}$, 
 and
 $ \xinverse{\epsilon}\indices{_{m}^{k'l'}} {\epsilon}\indices{^{m}_{kl}} = \delta\indices{^{k'}_{k}} \delta\indices{^{l'}_{l}}$. 
To get the speed-squared matrix, 
 we combine the above equations in the usual way 
 to derive a second order equation.
In terms of the $B$ field, 
 and for homogeneous material parameters,
 this is 
~
\begin{align}
  \partial_t^2 B^m
&=
  \partial_i 
    {\CSQ}\indices{^{imj}_l} 
    \partial_j
      B^l
.
\end{align}
Here the generalized (four index)
 speed-squared matrix ${\CSQ}\indices{^{imj}_l}$
 depends on antisymetrized versions
 of the permitivitty $\epsilon\indices{_{ij}}$
 and inverse permeability $\eta\indices{_{kl}}$; 
 and are segments (blocks) excised from the dual of the constitutive tensor 
 as used by Kinsler and McCall \cite{Kinsler-M-2014pra}\footnote{However, 
  note that although that paper used $\chi\indices{^{\mu\nu}_{\gamma\delta}}$
  for the dual of the usual 
  EM constitutive tensor $\chi\indices{^{\mu\nu\gamma\delta}}$, 
  it denoted this only by the alternate indexing, 
  not (as would often be done)
  by $\dualchi\indices{^{\mu\nu}_{\gamma\delta}}$.
Also,
 it is helpful to note that if
 the vector components of the usual electric displacement 
 $\Vec{D}$ are denoted $D_i$ for $i \in \{x,y,z\}$
 and are related to $\Vec{E}$ (or $E_i$)
 by the usual vector-calculus style expression 
~
\begin{align}
 D_i 
&=
  \varepsilon_{ix} E_x
 + \varepsilon_{iy} E_y
 + \varepsilon_{iz} E_z
,
\end{align}
 then the components of the $j$-th slice of 
 $\epsilon\indices{^j_{kl}}$ in \eqref{eqn-tensorEM-constitutives}
 are
~
\begin{align}
  \left[ \epsilon\indices{^j_{kl}} \right]
&=
  \frac{1}{2}
  \begin{bmatrix}
     0                 &  -\varepsilon_{jz}  & +\varepsilon_{jy} \\
    +\varepsilon_{jz}  &   0                 & -\varepsilon_{jx} \\
    -\varepsilon_{jy}  &  +\varepsilon_{jx}  &  0 \\
  \end{bmatrix}
.
\end{align}
A similar expression can be found for 
 the $j$-slices of $\eta\indices{^{kl}_{j}}$.}. 
They are 
~
\begin{align}
  \epsilon\indices{^m_{kl}} = \dualchi\indices{^{m0}_{kl}}
,
\qquad
\textrm{and}
\qquad
  \eta\indices{^{mn}_j} = \dualchi\indices{^{mn}_{0j}}
.
\label{eqn-tensorEM-constitutives}
\end{align}
The
 multi-polarization speed-squared matrix 
 combines these as
~
\begin{align}
  {\CSQ}\indices{^{klm}_j}
&=
  \begin{bmatrix}
    \epsilon\indices{^n_{kl}}
  \end{bmatrix}^{-1}
  \eta\indices{^{nm}_j}
=
    \xinverse{\epsilon}\indices{_n^{kl}}
  \eta\indices{^{nm}_j}
\quad
\rightarrow
\quad
  {\CSQ}\indices{^{km}}
  \delta\indices{^l_j}
,
\end{align}
 where the simpler form indicated by the arrow 
 assumes the typical case where
 the constitutive parameters provide 
 no cross-coupling between polarizations.
{Although this excludes many more
 general types of media, 
 it nevertheless includes most of those
 of relevance in transformation devices,
 which almost invariably are single polarization only --
 but see \cite{Zentgraf-VTLZ-2010am} for an exception.
Further, 
 although anisotropic dielectric media are typically birefringent, 
 those generated by
 transformation from (or to) isotropic media
 are not \cite{McCall-KT-2016jo-helimed}.}

\pseudosection{}
As a final emphasis as to the value of the 
``speed squared'' denomination of a T-device, 
 note that a determination of
 the water wave speed {(squared)} profile
 was the natural one to use when designing and building
 the Maxwell's Fishpond \cite{Kinsler-TTTK-2012ejp}.

The above shows us what material parameters will need
 to be engineered to control either waves or rays 
 for acoustic-like or EM scenarios.
{In all these three cases where we generate a speed-squared metric 
 for a physical system, 
 impedance does not appear in the final result.
Impedance is expressed as either the ratio of the two field components
 in a propagating wave 
 (but only \emph{one} field appears in a second order wave equation),
 or is the ratio of two constitutive quantities
 (but which \emph{only} appear as a product).}

The next piece of the puzzle is to 
 show what material parameters need to be manipulated
 to control diffusion processes.
Since an important case of diffusion is defined by
 the heat diffusion equation, 
 and it is one already considered
 by the T-Design community \cite{Guenneau-AV-2012oe},
 we will start there.\\

\pseudosection{Heat diffusion:~}
The diffusion equation 
 for a temperature distribution $u(\Vec{r},t)$,
 is often written in ordinary vector calculus notation as
$  \rho
  c_p
  \partial_t u
=
  \grad
  \cdot
  K
  \cdot
  \grad 
  u$
 for density $\rho$,           
 specific heat $c_p$,            
 and potentially anisotropic thermal conductivity matrix
 with components $K^{ij}$. 
The positioning of $K$ in this equation is the ``natural'' one, 
 and has not required any approximation involving slow/negligible
 spatial variation.

Here we first take a step back and write two first order equations
 in a style reminiscent of {both} 
 p-acoustics and the presentation of EM as above.
{Such rewriting is an important step in matching 
 the heat equation theory to the others,
 and as an enabler of our generalization approach.}
Although in the following we examine heat diffusion in particular, 
 the equations are equally applicable to other diffusion processes, 
 as long as the appropriate reinterpretations of the physical variables
 are made.
We therefore start with a conservation equation relating an energy density $h$ 
 and an energy flux ${v}^i$ which is 
~
\begin{align}
  \partial_t h
&=
 -
 \grad_i v^i
.
\label{eqn-pheat-dth}
\end{align}
We then rewrite the usual expression relating heat flux 
 ${V}^i$ to temperature profile $u$ which is 
 $V^i = K^{ij} \grad_j u$ using a `temperature impulse' ${W}^i$ as
~
\begin{align}
  \partial_t
  W^k
&=
 -
  \grad_i
  u^{ij}
.
\label{eqn-pheat-dtW}
\end{align}
Here the temperature profile 
 is allowed to be anisotropic, 
 which does not necessarily have a clear physical meaning; 
 typically we will expect $u^{ij}$ to be diagonal.

With the assumption that $u^{ij}$ is proportional to $h$, 
 but $W^i$ and $v^j$ are related by a first order differential equation, 
 we (can) write 
 the following constitutive relations (equations of state), 
~
\begin{align}
  u^{ij} &= \sigma^{ij} h
,
\label{eqn-pheat-u-h}
\\
  \partial_t
  W^k
&=
 -
  \gamma
  \beta^k_j
  v^j
.
\label{eqn-pheat-W-v}
\end{align}
Here $\sigma^{ij}$ plays the role of the inverse of the product $\rho c_p$.
Note that we will also need $\beta^k_j \xinverse{\beta}^j_l = \delta^k_l$.
Substitution of \eqref{eqn-pheat-dtW}
  into \eqref{eqn-pheat-dth} then gives us 
~
\begin{align}
  \partial_t
  h
&=
  \grad_l
  \left( \gamma^{-1} \xinverse{\beta}^l_j \right)   
  \grad_i
  u^{ij}
\quad
=
  \grad_l
  \left( \gamma^{-1} \xinverse{\beta}^l_j \right)   
  \grad_i
  \sigma^{ij}
  h
\label{eqn-diffusion-ph1}
\end{align}
This is just a diffusion equation for $h$, 
 and if $\sigma^{ij}$ is homogeneous 
 (or has negligible spatial variation),
 then
~
\begin{align}
  \partial_t
  h
&=
  \grad_l
  \left( \gamma^{-1} \xinverse{\beta}^l_j \right)   
  \sigma^{ij}
  \grad_i
  h
\quad
=
  \grad_i
  {\Dffsn}^{ij}
  \grad_j
  h
,
\label{eqn-diffusion-ph2}
\end{align}
 where in comparison to the original $K^{ij}$, $\rho$, $c_p$ quantities, 
~
\begin{align}
  {\Dffsn}\indices{^{ij}}
=
    G^l_j
  \sigma^{ij}
\quad
&\Leftrightarrow 
\quad
  {\Dffsn}\indices{^{ij}}
=
  K^{ij}/\left(\rho c_p\right)
.
\end{align}

Following this, 
 we now know how to relate the heat equation parameters
 to the diffusion, 
 and hence to the transformed metric, 
 just as for the wave and ray theories already described.
Note that other diffusion equations can also easily be recast
 into the form used above.

%
\section{Transformations}
\label{S-transformations}

From the preceeding section, 
 we can see that both (second order) wave and ray propagation, 
 and even diffusion processes
 can be packaged in a way dependent
 on the same mathematics;
 and that how that mathematics describes propagation 
 depends intimately on the metric.
{We must emphasize, 
 however, 
 that the significant gain in generalizing transformation design
 which we achieve here is not without cost --
 being that we neglect details present in more exact physical models.
But under appropriate approximations,
 we simply need to determine --}
 for a chosen transformation (deformation) --
 {the new metric,}
 given that we insist that energy transport 
 and ray trajectories (geodesics), 
 are shifted \emph{only} by that deformation.

\begin{figure}
  \begin{center}
    \includegraphics[width=\figwidthFull]{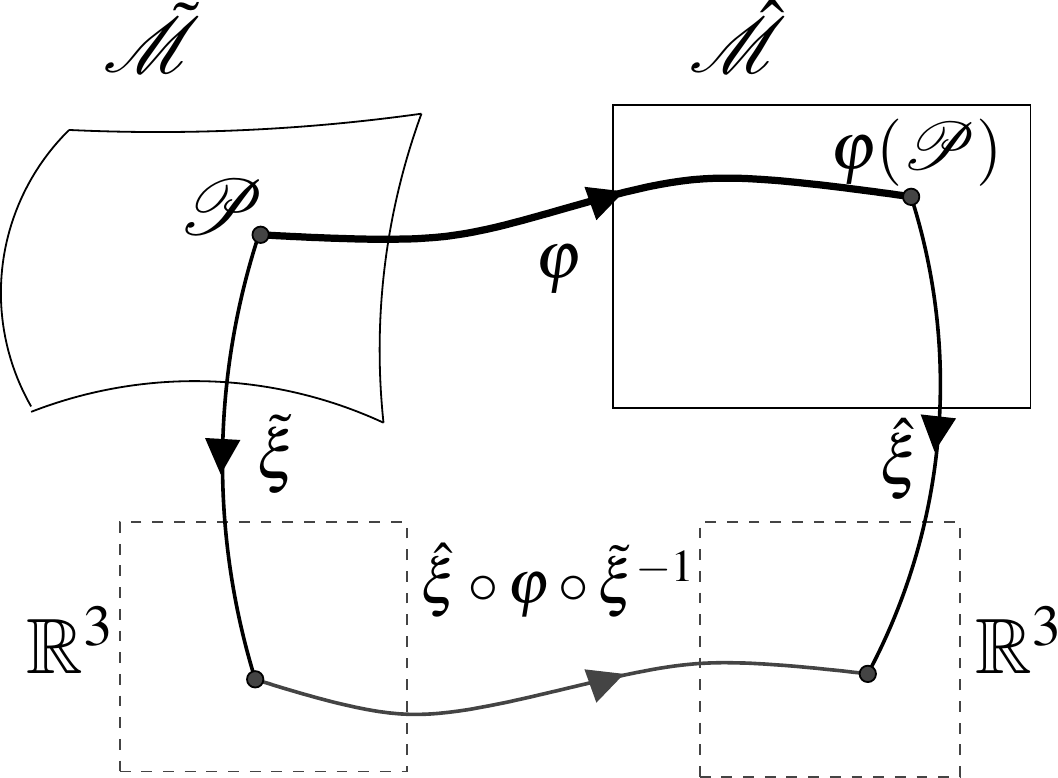}
  \end{center}
\caption[]
{ \label{fig-morphism}
(Diffeo)-morphism. 
Points $\Point$ in the device manifold $\manDev$
 are mapped to 
 points ${\Morph} (\Point)$ in the design manifold $\manDes$
 by the morphism (mapping) ${\Morph}$. 
The $\mReals^3$ coordinate representation of the morphism, 
 is $\Lambda^{-1} = \indesign{\xi} \circ {\Morph} \circ \indevice{\xi}^{-1}: 
 \mReals^3 \rightarrow \mReals^3$.
This means that the \emph{design} of the T-device is specified by ${\Morph}$,
 but that expression of that design
 in coordinate terms -- 
 the ``blueprint'' -- 
 is $\Lambda$. 
}
\end{figure}

%
\subsection{Metrics and coordinate transformations}

If we adhere to the traditional coordinate-based interpretation 
 of T-Design, 
 then we can transform the metric simply
 with the notional coordinate transform
 that we use to define our desired T-device.
Coordinate transformations of representations of tensors
 depend on the differential relationships 
 between the old and new coordinates, 
 i.e.
~
\begin{align}
  x^{\alpha'} &= f(x^\alpha)
\qquad
  \textrm{then} 
\qquad
  \Lambda\indices{^{\alpha'}_\alpha}
=
  \frac{\partial f(x^\alpha)}
       {\partial x^\alpha}
.
\label{eqn-xTf}
\end{align}

So a metric ${\gMetric}$ and its inverse
 re-represented in new primed coordinates
 would be 
~
\begin{align}
  {\gMetric}\indices{_{\alpha'\beta'}}
&=
  \Lambda\indices{^{\alpha}_{\alpha'}} 
  {\gMetric}\indices{_{\alpha\beta}}
  \Lambda\indices{^{\beta}_{\beta'}}
\label{eqn-gmetricLoTf}
,\\
  {\gMetric}\indices{^{\alpha'\beta'}}
&=
  \Lambda\indices{^{\alpha'}_{\alpha}} 
  {\gMetric}\indices{^{\alpha\beta}}
  \Lambda\indices{^{\beta'}_{\beta}}
.
\label{eqn-gmetricHiTf}
\end{align}

In the standard T-Design paradigm
 the coordinate transformation has the effect of changing
 our reference case geodesics into 
 new, useful, device geodesics; 
 then we need to adapt our material parameters from the reference values
 to those that give rise to the T-device metric ${\gMetric}_{\alpha'\beta'}$.

We have already seen that an expression of a metric ${\gMetric}_{\alpha\beta}$
 can be inverted to give speed-squared matrix ${\CSQ}^{ij}$; 
 thus the inverted T-device metric ${\gMetric}^{\alpha'\beta'}$
 tells us what the T-device speed-squared matrix ${{\CSQ}'}^{ij}$
 must be.
Our knowledge of whatever chosen physical system 
 we want to build the device using then tells us what 
 material properties are needed to achieve the necessary ${{\CSQ}'}^{ij}$
 and hence implement the T-device.

However, 
{if any physical idea must be expressible in a way 
 that is independent of coordinates; 
 how can any claimed ``coordinate transformation'' 
 hope to represent the design or specification
 of a new device?}
{Although 
 the coordinate transformation paradigm works
 from a purely practical standpoint,
 some coordinate transformations cannot be represented as a diffeomorphism --
 the transform from cartesian to polar coordinates is one such example.}
In what follows
 we present a more 
 {general and mathematically formal}
 method
 which encapsulates the 
 steps needed to rigorously implement the process of T-Design.

%
\subsection{Metric induced by a diffeomorphism}

The mathematical underpinning of all transformation theories
 is in fact not that of coordinate transformation, 
 but that of a \emph{morphism} $\Morph$
 that maps a point $\Point$ on a 
``device'' manifold $\manDev$, 
 to another point $\Morph (\Point)$ 
 on a reference (``design'') manifold $\manDes$,
 as seen in Fig. \ref{fig-morphism}.
A coordinate chart $\indesign{\xi}: \subDes \rightarrow \mReals^3$
 maps a point $\Point \in \subDes \subset \manDes$ to Euclidean space.
Mappings exist from the  manifolds $\manDev$ and $\manDes$
 into charts on $\mReals^3$,
 these are
 $\indesign{\xi}: \manDes \rightarrow \mReals^3$ at $\Point$,
 and $\indevice{\xi}: \manDev \rightarrow \mReals^3$ at $\Morph(\Point)$.
However,
 in T-Optics,
 $\Morph$ {enables us to prescribe} the electromagnetic medium
 in the device manifold $\manDev$ for {(typically)}
 a vacuum-like manifold $\manDes$,
 before any discussion of coordinates
 \cite{Thompson-CJ-2011jo,Thompson-CJ-2011jo-b}.

Our primary task here is to
 work out how to specify\footnote{A summary
 of the mathematical details associated with the transformation of metrics
 can be seen in the appendix.}
 the device metric ${\indevice{\gMetric}}_{ij}$,
 as induced by $\Morph$,
 in terms of 
 the reference/design metric ${\indesign{\gMetric}}_{ij}$.
Often ${\indesign{\gMetric}}_{ij}$ will just be a flat Minkowski metric $\eta_{ij}$, 
 but not always.
The device metric $\indevice{\gMetric}$ is related to the design metric $\indesign{\gMetric}$ by
~
\begin{align}
  {\indevice{\gMetric}}_{\Point} (X_\Point,Y_\Point) 
&=
  \left[
    \Morph^{\pstar}
   {\indesign{\gMetric}}
  \right] (X_\Point,Y_\Point) 
=
 {\indesign{\gMetric}}
 ( \Morph_{\pstar} X_\Point,  \Morph_{\pstar} Y_\Point) 
,
\label{eqn-g-transform}
\end{align}
 where we have to (either)
 \emph{pullback} the metric $\indesign{\gMetric}$,
 or \emph{pushforward} its arguments $X_\Point$, $Y_\Point$ 
 using the diffeomorphism $\Morph$.
With some thought, 
 we can see that this should be the expected behaviour:
 what we usually are typically trying to achieve is to make
 interesting trajectories into the ``new normal''.
E.g. we make a cloak-like device structure in the laboratory, 
 with specified cloak-like wave or ray paths, 
 into an actual cloak by insisting its properties
 are such that those paths look to the outside world
 as if they were in an unremarkable piece of vacuum.
I.e., 
 we are trying to make a device (cloak) manifold $\manDev$
 look like the design (vacuum) manifold $\manDes$, 
 and not the other way around. %
In contrast, 
 the traditional transformation proceedure
 acts like an active transformation,
 and in our morphism picture is
 from the design space 
 to the device space.
As a consequence 
 the traditional picture's basic operation
 is specified by $\Morph^{-1}$ rather than $\Morph$, 
 and in our notation would be written
~
\begin{align}
  \Lambda
=
  \indevice{\xi} \circ \Morph^{-1} \circ \indesign{\xi}^{-1} 
&:
  \mReals^3 \rightarrow \mReals^3
.
\label{eqn-transfomorphism}
\end{align}
This makes it clear
 that the traditional picture 
 is an active transformation\footnote{Note the distinction
  between an active transformation which changes the system,
 and therefore \emph{cannot} be considered as a coordinate transformation;
  and a passive transformation which 
  only changes the coordinate representation,
  i.e. which \emph{can} be considered as a coordinate transformation.}.

As we saw in the previous section,
 for our purposes the inverse of the metric, 
 which is related to our speed squared matrix $\CSQ$, 
 is more useful.
In mathematical terminology, 
 this is known as the co-metric, 
 which we will denote ${\gCoMetric}$.
The co-metric version of \eqref{eqn-g-transform}
 uses the pushforward  rather than the pullback,
 and is
~
\begin{align}
  {\indesign{\gCoMetric}}_{\Morph_{\pstar}\Point} 
  (\Morph_{\pstar} X_\Point, \Morph_{\pstar}Y_\Point) 
&=
  \left[
    \Morph_{\pstar}
   {\indevice{\mathfrak{\gMetric}}}
  \right] 
  (\Morph_{\pstar} X_\Point, \Morph_{\pstar}Y_\Point) 
=
 {\indevice{\gCoMetric}}
 ( X_\Point,  Y_\Point) 
,
\label{eqn-co-g-transform}
\end{align}

Although in a mathematical sense, 
 this has defined everything we need,
 for practical calculations a matrix notation is more  convenient.
The first step in achieving this
 is to write down \eqref{eqn-g-transform} and \eqref{eqn-co-g-transform}
 in an indexed notation;
 after which a choice of coordinates leads us to the relevant matrix form.
Notably, 
 \eqref{eqn-g-transform} can be written
~
\begin{align}
  {\indevice{\gMetric}}\indices{_{\indevice{\mu}\indevice{\nu}}}
&=
  (\Morph^{\pstar})\indices{^{\indesign{\alpha}}_{\indevice{\mu}}}
  (\Morph^{\pstar})\indices{^{\indesign{\beta}}_{\indevice{\nu}}}
  ~
  {\indesign{\gMetric}}\indices{_{\indesign{\alpha}\indesign{\beta}}}
,
\label{eqn-inducedmetric-indexed1}
\end{align}
 where
\begin{align}
    (\Morph^{\pstar})\indices{^{\indesign{\beta}}_{\indevice{\nu}}}
=
  \left.
    \frac{\partial y^{\indesign{\beta}}}
         {\partial x^{\indevice{\nu}}}
  \right|_{\Point}
.
\label{eqn-morphismindexed}
\end{align}

\def\MorphI{(\Morph^{-1})}

Similarly, 
 we can do the same for the co-metric ${\gCoMetric}$.
Since it has raised (and not lowered)
 indices, 
 this distinguishes the the co-metric from the metric, 
 and so we can replace the ${\gCoMetric}$ with an ordinary ${\gMetric}$,
 which matches with the notation used for the inverse metric 
 in previous sections.
We then can write 
~
\begin{align}
  {\indesign{{\gMetric}}}\indices{^{\indesign{\alpha}\indesign{\beta}}}
&=
  (\Morph_{\pstar})\indices{^{\indesign{\alpha}}_{\indevice{\mu}}}
  (\Morph_{\pstar})\indices{^{\indesign{\beta}}_{\indevice{\nu}}}
  ~
  {\indevice{{\gMetric}}}\indices{^{\indevice{\mu}\indevice{\nu}}}
,
\label{eqn-co-inducedmetric-indexed1}
\\
  {\indevice{{\gMetric}}}\indices{^{\indevice{\mu}\indevice{\nu}}}
&=
  (\MorphI_{\pstar})\indices{^{\indevice{\mu}}_{\indesign{\alpha}}}
  (\MorphI_{\pstar})\indices{^{\indevice{\nu}}_{\indesign{\beta}}}
  ~
  {\indesign{{\gMetric}}}\indices{^{\indesign{\alpha}\indesign{\beta}}}
,
\label{eqn-co-inducedmetric-indexed2}
\end{align}
 where
\begin{align}
    (\MorphI_{\pstar})\indices{^{\indevice{\nu}}_{\indesign{\beta}}}
=
  \left.
    \frac{\partial x^{\indevice{\nu}}}
         {\partial y^{\indesign{\beta}}}
  \right|_{\Morph_\pstar\Point}
.
\label{eqn-co-morphismindexed}
\end{align}

Notably, 
 if we were to replace the ``$(\Morph)$'' notation 
 with a coordinate-transform mimicking ``$\Lambda$'', 
 then the operations done here, 
 in this more sophisticated 
 pushforward/ pullback
 would match the use in the so-called ``coordinate transform'' approach 
 (see e.g. \cite{Kinsler-M-2014pra}); 
 albeit with the significant advantage
 of being 
 {better} 
 mathematically and physically {motivated}.
Moreover, there is a further distinction to be emphasised. 
In two dimensions,
 for example,
 one can have charts $\phi_1: \manM \rightarrow \mReals$
 and $\phi_2: \manM \rightarrow \mReals \otimes \mCircles$
 and the transition map $\phi_2 \circ \phi_1^{-1}$
 describing the transformation from Cartesian to polar coordinates. 
This coordinate transformation and the associated $\Lambda$ matrices
 can be readily inserted into
 \eqref{eqn-gmetricLoTf} or \eqref{eqn-gmetricHiTf}
 but this does not produce a T-device. 
Morphing points on a manifold,
 on the other hand,
 one can use a single coordinate chart
 $\indesign{\xi}: \manM \rightarrow \manifold{U}$
 and the transition function $\indesign{\xi} \circ \Morph \circ \indesign{\xi}^{-1}$
 (cf. Fig. \ref{fig-morphism})
 to reach a well-defined,
 and useful,
 T-device,
 by morphing the metric according to \eqref{eqn-inducedmetric-indexed1}.
Also note that if the `morphism'
 is to leave points on the manifold unchanged,
 then the $\Morph$ matrices are the identity,
 whereas the $\Lambda$ matrices
 associated with coordinate transformations need not be.

%
\subsection{Transformations, Morphisms}

It is critical to notice at this point that if we specify the morphism
 in terms of how one coordinate point is moved to another, 
 then we are only adjusting distances; 
 i.e. only adjusting the effective metric.
This is because we have explicitly separated the general step 
 of specifying the device by means of a morphism,
 from its chosen implementation in a particular physical system.
As a result our design/morphism 
 is only targetted
 at the speed-squared matrix ${\CSQ}\indices{^{ij}}$ 
 (i.e. the inverse metric), 
 and nothing else.
It cannot directly specify
 how physical properties such as field values, 
 material parameters,
 ratios, 
 or impedances have been affected, 
 since they are still undecided.

However, 
 once we \emph{have} taken the additional step of specifying the 
 physical system (e.g. EM),
 we can use the morphism to tell us how physical properties will be changed, 
 and how much freedom there might be.
For example, 
 choosing the usual kappa medium assumption 
 of ``${\epsilon}={\mu}$''
 in an electromagnetic scenario has implications 
 for impedance \cite{Kinsler-2015zloak}, 
 but such an identification has nothing to do with the morphism itself, 
 which says nothing about the $\epsilon:\mu$ ratio.
Of course, 
 given a specified physical system, 
 it is possible to apply T-design techniques to 
 transform fields and/or other properties independently
 \cite{Liu-J-2015prl,Liu-LL-2013prl}.

As described above, 
 here we make an identification between
 the metric and the material properties, 
 in the same way as introduced to T-Optics
 by Leonhardt and Philbin \cite{Leonhardt-P-2006njp}.
This is essentially a ``coordinate redefinition'' step, 
 which results from the choosing of a secondary map 
 \cite{Fathi-T-2016prd}.

%
\section{Examples}
\label{S-examples}

As noted above, 
 typically T-devices are described with a ``transformation'' narrative, 
 where we talk of transforming an unremarkable reference space
 into an interesting device space.
Hence the typical description of a cloaking transformation 
 being that of
 a point in a flat space being
 expanded and pushed outwards to form a disk, 
 and where (outwardly)
 the inside of that disk (``core'') region is invisible.

In the more rigorous morphism language
 we instead represent the deformation that takes the 
 device (or ``laboratory'') space manifold ($\manDev$), 
 with its missing disk, 
 and alters the metric on that manifold so as to ``pull it inwards'.
As a result $\manDev$ then acts as if it were
 like a design (i.e. apparent, or ``reference'') manifold ($\manDes$)
 only missing a single point.
This is the reverse narrative of the (usual) transformation one, 
 and the mathematical and physical reasons for this
 were described in the previous section.

However, 
 the reason why the usual transformation narrative
 is not without its uses
 is that non-trival reference manifolds
 might have metrics and geodesics with all kinds of interesting properties, 
 involving ones that have foci, caustics, or that form loops.
And whatever the exotic properties of our T-device
 might be in re-presenting the physical reality to an observer, 
 it must be capable of being mapped onto that apparent manifold.
By starting a design process with the intended (design) behaviour, 
 and morphing (by pullback) to the device behaviour,
 we can guarantee that our aim is achievable, 
 at least in principle.
This specification means that the morphism should also
 have differentiable inverse, 
 i.e. be a diffeomorphism.

In the examples below, 
 the necessary device metric $\indevice{\gMetric}_{ij}$ can be calculated from 
 the design metric $\indesign{\gMetric}_{ij}$
 using the components $(\Morph)\indices{^i_l}$
 of the design morphism $\Morph$.
Using square brackets to indicate a matrix-like representation, 
 we find that 
~
\begin{align}
  \begin{bmatrix}
    \indevice{\gMetric}_{ij}
  \end{bmatrix}
&=
  \begin{bmatrix}
    (\Morph^{\pstar})\indices{^l_i} 
  \end{bmatrix}
  \begin{bmatrix}
    \indesign{\gMetric}_{lm}
  \end{bmatrix}
  \begin{bmatrix}
    (\Morph^{\pstar})\indices{^m_j}
  \end{bmatrix}^T
,
\\
  \begin{bmatrix}
    \indevice{\gMetric}^{ij}
  \end{bmatrix}
&=
  \begin{bmatrix}
    (\MorphI_{\pstar})\indices{^i_l}
  \end{bmatrix}
  \begin{bmatrix}
    \indesign{\gMetric}^{lm}
  \end{bmatrix}
  \begin{bmatrix}
    (\MorphI_{\pstar})\indices{^j_m}
  \end{bmatrix}^T
=
  \begin{bmatrix}
    \indevice{\CSQ}^{ij}
  \end{bmatrix}
.
\end{align}
Here the second line represents the next,
 more pragmatic step,
 where the inverse device metric is used to generate the 
 speed-squared matrix $[\indevice{\CSQ}^{ij}]$ that we need
 to engineer using the relevant material properties.
Note that this transformation also uses the $\Morph_{\pstar}$ pushforward
 form of the inverseof the morphism,
 i.e. $\Morph^{-1}$.
In the examples that follow, 
 we will use the phrase ``non-trivial''
 to describe any diffeomorphism components $(\MorphI_{\pstar})\indices{^i_l}$
 that differ from the identity transformation
 value of $\delta\indices{^i_l}$.
For example,
 if we chose to restrict ourselves to a cylindrical geometry,
 with only radial transformations,
 the only non-trival components will be radial ones.

Further, 
 we also show only ray examples, 
 because strictly speaking only in the ray limit is the metric approach exact.
In any case, 
 the literature is already full of wave-cloak pictures
 for various degrees of approximation in the underlying model.
A careful analysis demonstrating the effects
 of the neglected impedance terms and
 {possibly non-trivial properties of the underlying space}
 is no simple matter, 
 and will be addressed elsewhere \cite{Kinsler-2015zloak,Kinsler-2015gloak}.

%
\subsection{Cylindrical cloak}
\label{S-examples-ccylinder}

The cylindrical cloak first introduced by Pendry et al \cite{Pendry-SS-2006sci}
 is the most famous T-device.
Its design is usually expressed as expanding a central point into a disk 
 (or \emph{``core''}) $r=R$
 in diameter, 
 while compressing its \emph{``halo''} --
 the  space between the point and the outer rim at $r=S$
 accordingly.
It therefore preserves the effective distance between inner and outer radii
 as being the same as (just) the outer radius.
It is usually written in cylindrical coordinates, 
 and a general form allowing for a variety of radial transformations is
~
\begin{align}
  \indesign{r} = {f}(\indevice{r}),
  \quad \indesign{\theta} = \indevice{\theta},
  \quad \indesign{z}=\indevice{z}
,
\label{eqn-eg-cylcloak-coords}
\end{align}
 where $f(\indevice{r})$ is some suitably well behaved function
 increasing from $f(R)=0$ to $f(S)=S$;
 it is the derivative of this $f$ which 
 specifies
 the only non-trivial morphism component  $(\Morph^{\pstar})\indices{^r_r}$.

In the original proposal \cite{Pendry-SS-2006sci},
 $f$ was simply the linear 
~
\begin{align}
  f(\indevice{r}) = \frac{S}{S-R}\left(\indevice{r}-R\right)
.
\end{align}

Here $\indevice{r}, \indevice{\theta}, \indevice{z}$
 are the device coordinates
 where waves or rays are confined to $\indevice{r}>R$,
 and so hiding points where $\indevice{r}<R$.
Otherwise, 
 the coordinates $\indesign{r}, \indesign{\theta}, \indesign{z}$
 span the design space
 where waves or rays are allowed at any $\indesign{r}>0$.

Let us assume for additional generality that we want the 
 design spatial metric
 (i.e. the apparent metric of our device)
 to have 
 independent radial, angular, and axial refractive index profiles
 $n(\indesign{r})$, $m(\indesign{\theta})$, $s(\indesign{z})$.
This allows us, 
 for example,
 to consider adding a cloak to a 2D Maxwell's fisheye device
 \cite{Sarbort-T-2012jo,Luneberg-MTO}, 
 where $m(\indesign{\theta})=1$, $s(\indesign{z})=1$
 and $n(\indesign{r}) = n_0 / [1+(r/r_0)^2]$.
Our design metric is then 
~
\begin{align}
  {d\indesign{S}}^2 
&=
  n^2(\indesign{r}) 
  {d\indesign{r}}^2
 +
  m^2(\indesign{\theta}) \indesign{r}^2 
  {d\indesign{\theta}}^2
 +
  s^2(\indesign{z}) 
  {d\indesign{z}}^2
,
\\
  \textrm{or}
\qquad
  \indesign{\gMetric}\indices{_{ij}} 
&= 
  \begin{bmatrix}
   n^2     & ~~0~~  & ~~0~~ \\
  ~~0~~    &  m^2   & ~~0~~ \\
    0      &   0    &  s^2
  \end{bmatrix}
.
\end{align}
This means that whilst an outside observer 
 with no reason to make complicating assumptions
 would presume geodesics which match those in the design spatial metric, 
 that region can have properties that differ
 according to some morphism $\Morph$.
A morphism $\Morph$
 based on the transformation $f$ from \eqref{eqn-eg-cylcloak-coords}
 has the important (non trivial) component 
 $(\Morph_A)\indices{^r_r} = f' = \partial f/ \partial \indevice{r}$.
With this, 
 it hides the core region
 as a result of generating  
 the required device spatial metric of
~
\begin{align}
  {d\indevice{S}}^2 
&=
  n^2(f(\indevice{r}))
  \left[ \frac{d f(\indevice{r})}{d \indevice{r}} \right]^2 
  {d\indevice{r}}^2 
 +
  m^2(\indevice{\theta})
  \left[ \frac{f(\indevice{r})}{\indevice{r}} \right]^2 \indevice{r}^2 {d\indevice{\theta}}^2 
\nonumber
\\
& \qquad\qquad\qquad\qquad\qquad\qquad
 +
  s^2(\indevice{z})
  {d\indevice{z}}^2
,
\end{align}
 which, 
 as should be expected, 
 looks like (is) the same result as that obtained 
 by the misleadingly named
 ``coordinate transformation''
 approaches \cite{Cummer-LC-2009jap,Cai-CKSM-2007apl}
 based on equations like \eqref{eqn-eg-cylcloak-coords}.
This interval-style ${d{S}}^2 $ metric 
 can also be written in a matrix-like form, 
 with $n \equiv n(f(\indevice{r}))$, 
 $m \equiv m(\indevice{\theta})$, 
 $s \equiv s(\indevice{z})$, 
 i.e. 
 ~
\begin{align}
  \indevice{\gMetric}\indices{_{ij}} 
&= 
  \begin{bmatrix}
    n^2 {f'}^{2}    & 0          & ~~0~~ \\
    0           & m^2 f^{2}/\indevice{r}^2  & ~~0~~ \\
    0           &  0         & ~~s^2~~
  \end{bmatrix}
\end{align}

As described above 
 we can convert --
 by a simple inversion --
 this T-device metric into the corresponding speed-squared matrix,
 i.e.
~
\begin{align}
  \indevice{\CSQ}\indices{^{ij}} 
&= \indevice{\gMetric}\indices{^{ij}} 
= \begin{bmatrix}\indevice{\gMetric}\indices{_{ij}}\end{bmatrix}^{-1}
\end{align}

\begin{figure}
  \begin{center}
    \includegraphics[width=\figwidthFull]{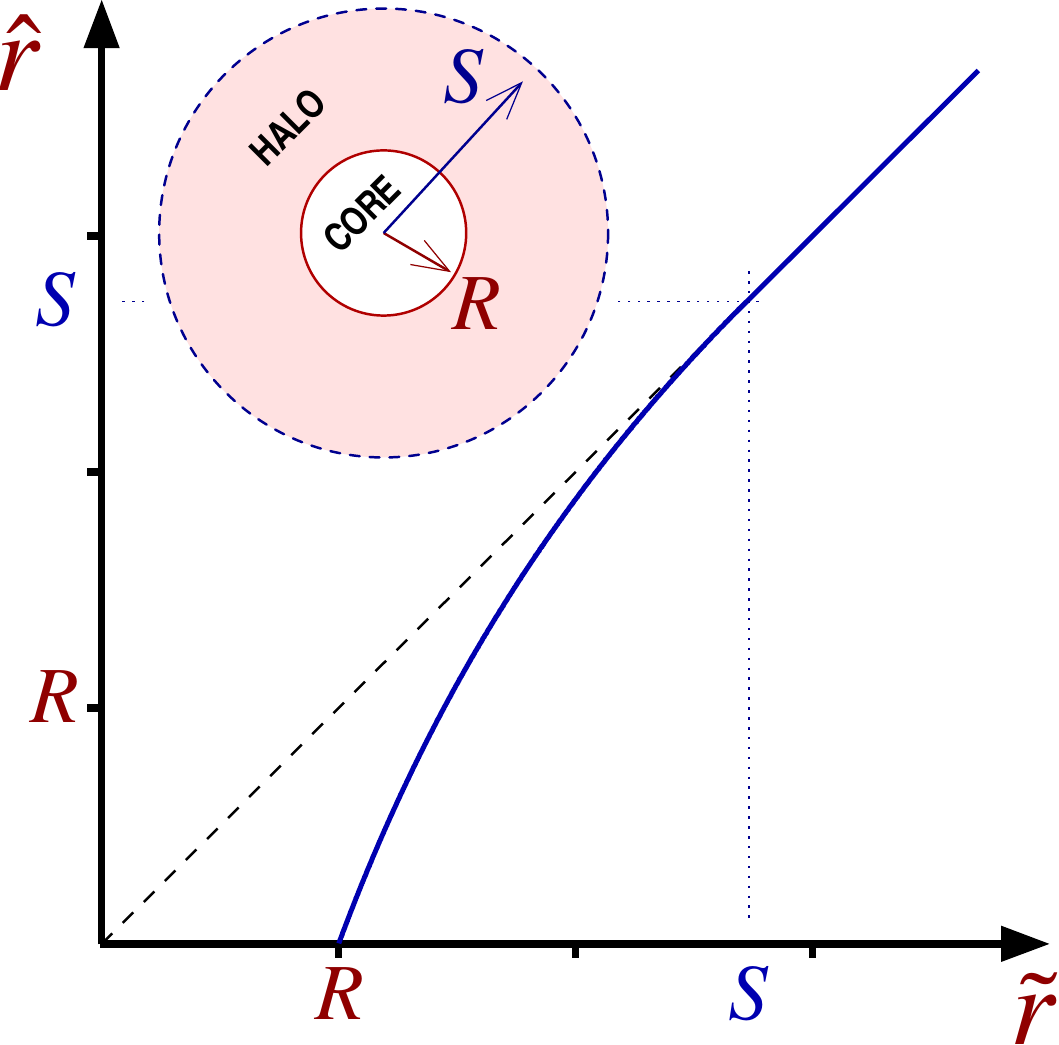}
  \end{center}
\caption{Radial cloak
 based on the logarithmic function
 given in \eqref{eqn-radialcloak-logarithmic}.
One nice feature of this cloaking function is that 
 it has both index and gradient matching 
 at the outer boundary $S=eR$ 
 of the halo. 
}
\label{fig-radcloak}
\end{figure}

We might, 
 for example, 
 {use an alternate}
 radial cloak using the logarithm function
 \cite{Ma-CYL-2009piers}
 so that it could be 
 {more smoothly}
 matched 
 than the original (linear) radial cloak \cite{Pendry-SS-2006sci}, 
 at its outer boundary.
The log radial cloak 
 is designed using
~
\begin{align}
  F(\indevice{r}) 
&=
  S \log \left[ \indevice{r}/R \right]
,
\quad \textrm{so that} \quad
  F'(\indevice{r}) 
&=
  S /\indevice{r}
.
\label{eqn-radialcloak-logarithmic}
\end{align}
To work,
 this log radial cloak requires a fixed core-to-halo 
 ratio so that $S=eR$.
The mapping between $\indevice{r}$ and $\indesign{r}$
 is shown on fig. \ref{fig-radcloak}.
The disadvantage of this design is that there is a stronger gradient
 at its interface with the core than with the original; 
 for particular experimental implementations, 
 this disadvantage may outweigh the benefits of its 
 {smoother}
 matching at the outer boundary.
This is because cloak performance can be strongly affected 
 by imperfect implementation of the inner (core) boundary, 
 although this would not be relevant in the near-miss case of a narrow beam 
 that only passes through the outer part of the cloak halo.

\begin{figure}
  \begin{center}
    \includegraphics[width=\figwidthHalf,angle=0]{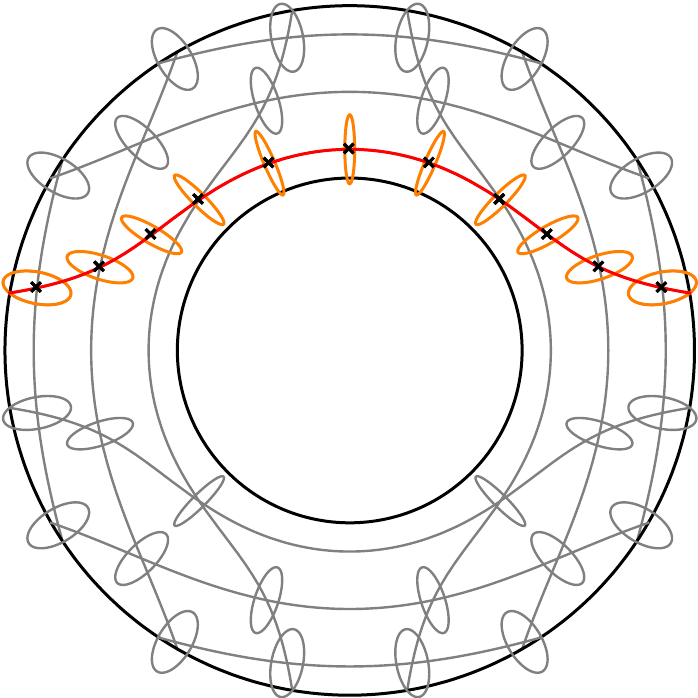}
    \includegraphics[width=\figwidthHalf,angle=0]{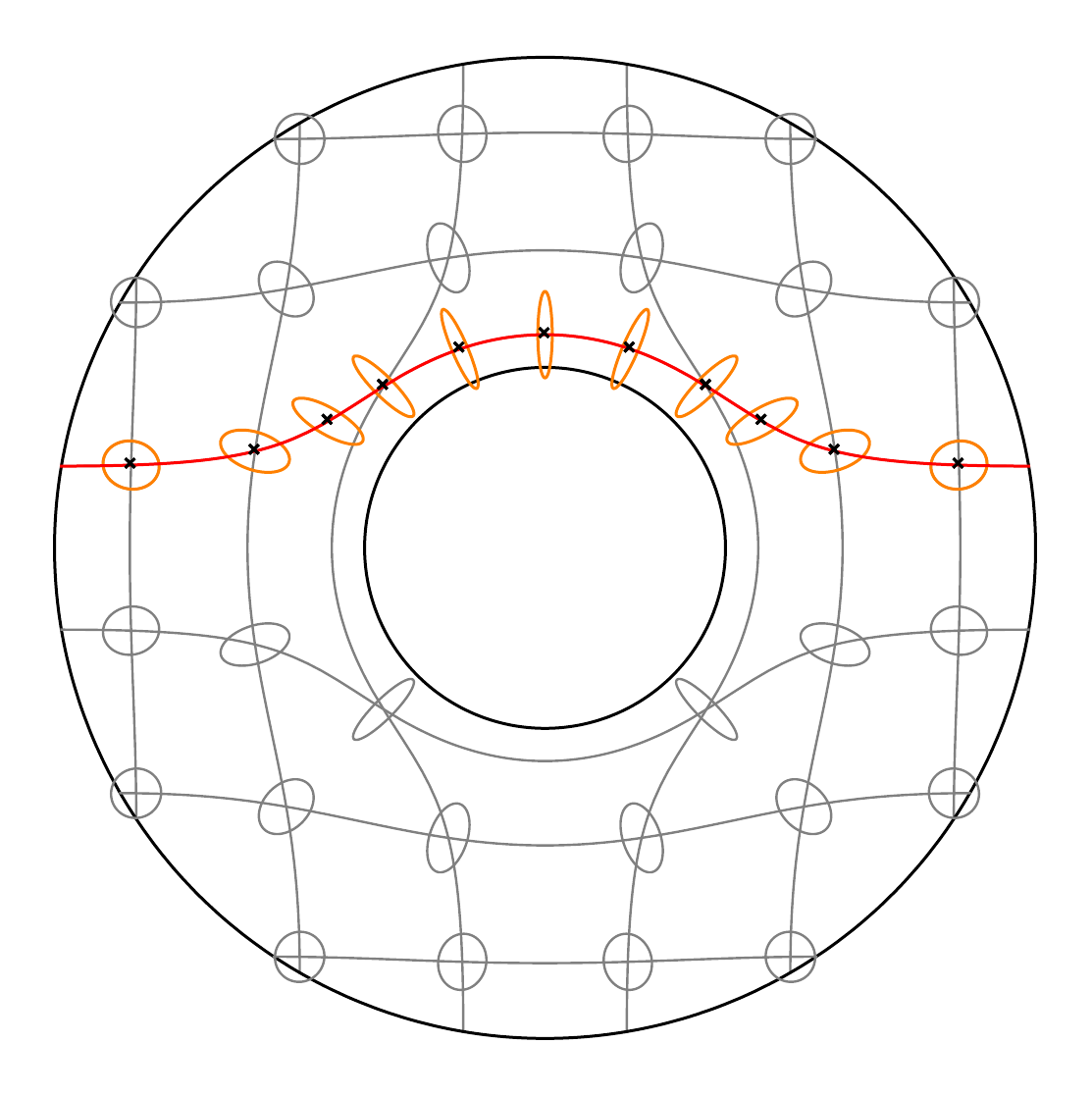}
  \end{center}
\caption{Comparison of the linear-radial 
 \cite{Pendry-SS-2006sci} and log-radial 
 cloaking transforms
 designed to match a flat space.
Sample of ray trajectories and 
 refractive index ellipses are shown.
Figures provided by R.D. Topf.
}
\label{fig-cfradial}
\end{figure}

So for an EM wave in this T-device cloak where
 we choose the case where
 the electric polarization is not aligned along $\theta$, 
 we find that 
 $\epsilon_{z}, \epsilon_{r}, \mu_r, \mu_\theta$
 are the controlling constitutive parameters.
Therefore in cylindrical coordinates
 we have
 ~
\begin{align}
  \indevice{\CSQ}\indices{^{ij}} 
= 
  \begin{bmatrix}
    \epsilon_{z}\mu_\theta & 0                 & 0 \\
    0                      & \epsilon_z \mu_r  & 0 \\
    0                      & 0                 & \epsilon_r \mu_\theta
  \end{bmatrix}^{-1}
= 
  \begin{bmatrix}
    1/n^2 {F'}^{2} & 0                          & ~~0~~ \\
    0              & \indevice{r}^2 / m^2 F^{2} & ~~0~~ \\
    0              &  0                         & ~~1/s^2~~
  \end{bmatrix}
.
\end{align}
Alternatively, 
 we might have chosen the complementary case where
 the magnetic field is not aligned along $\theta$.

For a scalar p-acoustics wave in the same cloaking T-device
 there is \emph{no} scope for a choice of polarizations, 
 unlike the EM case above.
In cylindrical coordinates we have
 a unique specification for the constitutive parameters that is
~
\begin{align}
  \indevice{\CSQ}\indices{^{ij}} 
= 
  \kappa_o
  \begin{bmatrix}
    \rho_{rr} & 0                   & 0 \\
    0         & \rho_{\theta\theta} & 0 \\
    0         & 0                   & \rho_{zz}
  \end{bmatrix}^{-1}
= 
  \begin{bmatrix}
    1/n^2 {F'}^{2} & 0                         & ~~0~~ \\
    0              & \indevice{r}^2 /m^2 F^{2} & ~~0~~ \\
    0              &  0                        & ~~1/s^2~~
  \end{bmatrix}
.
\end{align}

A comparison of the 
 EM and scalar p-acoustics cases
 is instructive.
In EM, 
 for each of the two possible field polarizations
 we can implement the three necessary cloak parameters
 $\indevice{\CSQ}^{rr}, \indevice{\CSQ}^{\theta\theta}, \indevice{\CSQ}^{zz}$
 using up to 
 six constitutive parameters, 
 i.e. we have three ``spare'' degrees of freedom.
In contrast, 
 the p-acoustic version has only four constitutive parameters
 with which to make up the three cloak parameters, 
 i.e. only one degree of freedom:
 we could e.g. leave $\kappa_o$ untouched and engineer each of 
 $\rho_{rr}, \rho_{\theta\theta}, \rho_{zz}$.
This significantly impacts our ability to fine-tune the p-acoustic cloak 
 in response to technological constraints.

A pentamode p-acoustics wave
 has additional constitutive freedom compared to the scalar version,
 since $\kappa$ is now matrix-like.
In cylindrical coordinates, 
 the same cloaking T-device has 
 a specification for the constitutive parameters
 based on diagonal $\kappa^{ij}$ and ${\rho}\indices{^k_l}$ properties is
~
\begin{align}
  \indevice{\CSQ}\indices{^{ij}} 
= 
  \begin{bmatrix}
    \frac{\kappa_{rr}}{\rho_{rr}} & 0                   & 0 \\
    0         & \frac{\kappa_{\theta\theta}}{\rho_{\theta\theta}} & 0 \\
    0         & 0                   & \frac{\kappa_{zz}}{\rho_{zz}}
  \end{bmatrix}
= 
  \begin{bmatrix}
    1/n^2 {F'}^{2} & 0                         & ~~0~~ \\
    0              & \indevice{r}^2 /m^2 F^{2} & ~~0~~ \\
    0              &  0                        & ~~1/s^2~~
  \end{bmatrix}
.
\end{align}
Note that 
 $\kappa^{nj}$ and $\xinverse{\rho}\indices{_n^i}$ need not be diagonal
 as long as their product $\indevice{\CSQ}\indices{^{ij}}$ is.

However, 
 despite caveats regarding impedance matching, 
 \emph{the wave or ray ``steering'' performance of these implementations
 is as identical as the identical metrics upon which they are based}; 
 only their scattering properties are different.

In the case of heat diffusion, 
 we can just implement the $\indevice{\CSQ}\indices{^{ij}}$
 directly as the diffusion matrix 
~
\begin{align}
  \indevice{\Dffsn}\indices{^{ij}} 
= 
  \frac{1}{\rho c_p} 
  \left[ {K}\indices{^{ij}} \right]
=
  \begin{bmatrix}
    1/n^2 {F'}^{2} & 0                         & ~~0~~ \\
    0              & \indevice{r}^2 /m^2 F^{2} & ~~0~~ \\
    0              &  0                        & ~~1/s^2~~
  \end{bmatrix}
,
\end{align}
 with constant density $\rho$ and 
 specific heat $c_p$,   
 and an anistropic 
 thermal conductivity $K\indices{^{ij}}$.  

%
\subsection{Cloak on a sphere}
\label{S-examples-csphere}

Here our design space is a 2D spherical surface of radius $R$
 which is most naturally expressed in spherical polar coordinates.
Allowing for independent angular indices  
 $T(\indesign{\theta})$, $P(\indesign{\phi})$, 
 this has a design spatial metric which is
~
\begin{align}
  d\indesign{S}^2 
&=
  \left[ T^2(\indesign{\theta}) \right]
  R^2 d\indesign{\theta}^2
 +
  \left[ P^2(\indesign{\phi}) \right]
  R^2 \sin^2(\indesign{\theta}) d\indesign{\phi}^2
,
\end{align}
 with the variation in polar angle $\indesign{\theta}$ 
  denoting lines of ``longitude'', 
 and variation in the azimuthal angle $\indesign{\phi}$ 
  being latitude.
The sensible choice is to orient the coordinates 
 so that both the missing point in the design (target) space, 
 and the missing spherical cap in the device (laboratory) space,
 are centered on the pole.
In this case we leave the azimuthal $\phi$ untouched
 so that $\indevice{\phi}=\indesign{\phi}$,
 but offset and rescale the polar angle $\indesign{\theta}$ so that  
~
\begin{align}
  \indesign{\theta} =  f(\indevice{\theta})
 \quad
 \textrm{and}
 \qquad
  \indevice{\theta} = f^{-1}(\indesign{\theta}) = F(\indesign{\theta})
.
\end{align}
Thus the morphed device metric, 
 with $(\Morph_B)\indices{^\theta_\theta} = f'(\indevice{\theta}) = df(\indevice{\theta})/d{\indevice{\theta}}$, 
 is
~
\begin{align}
  d{S}^2 
&=
  \left[
    T(f(\indevice{\theta}))
    ~
    {f'(\indevice{\theta})}
  \right]^2
  ~
  R^2 d{\indevice{\theta}}^2 
\nonumber
\\
&\qquad\qquad
 +
  \left[
    P(\indevice{\phi}) 
    \frac{\sin \left( f(\indevice{\theta}) \right)}
         {\sin (\indevice{\theta})}
  \right]^2
  ~
  R^2 
  \sin^2 (\indevice{\theta})
  d\phi^2
.
\end{align}
 This has been written 
 {so as to separate
 the part of the new T-device metric
 which encodes the necessary constitutive properties,
 (which have been put in square brackets)
 from that which encodes the spherical geometry}.
Our T-device needs the longitude component of rays (or waves) 
 to see an index a factor of $f'(\indevice{\theta})$
 larger than the background; 
 whereas the latitude component
 needs to see a different index depending on how close to the cloak core
 (at $\theta_0$) they are.

\begin{figure}
  \begin{center}
    \includegraphics[width=\figwidthFull,angle=0]{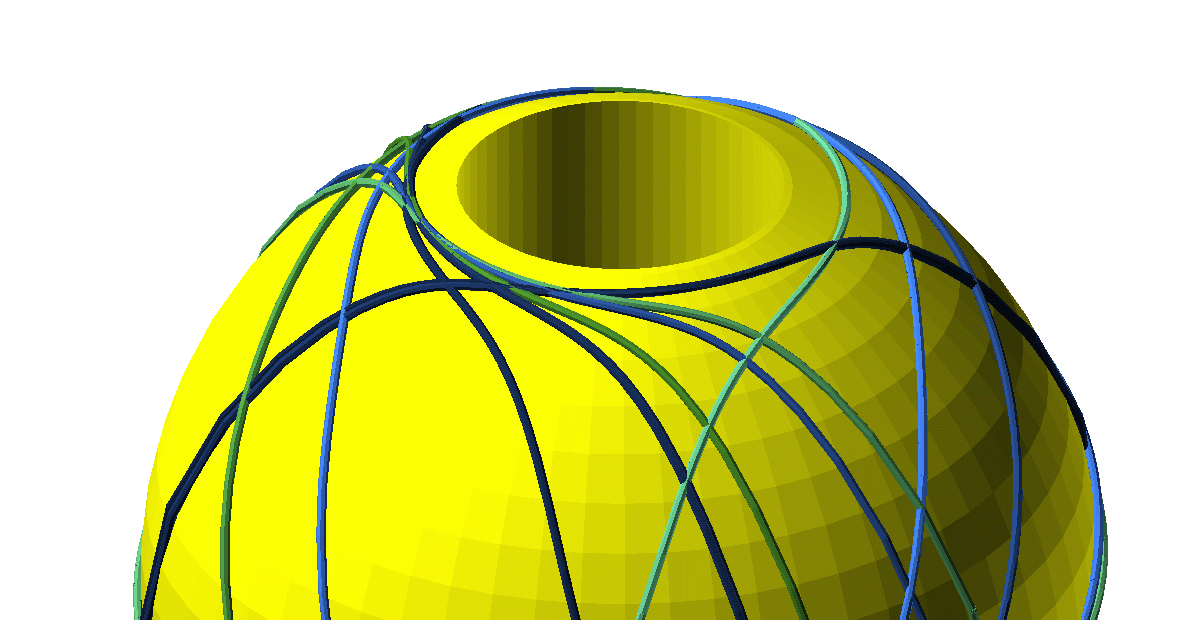} 
  \end{center}
\caption{Radial cloak
 on a spherical surface, 
 again based on a logarithmic function.
A random selection ray paths are shown, 
 all being forced by the cloak to 
 avoid the core region which extends up to $20^\circ$ from the north pole, 
 whilst also returning smoothly to their expected ``great circle'' paths 
 outside the cloak halo at $e \times 20^\circ \simeq 54^\circ$.
}
\label{fig-spherecloak}
\end{figure}

\begin{figure}
  \begin{center}
    \includegraphics[width=\figwidthFull,angle=0]{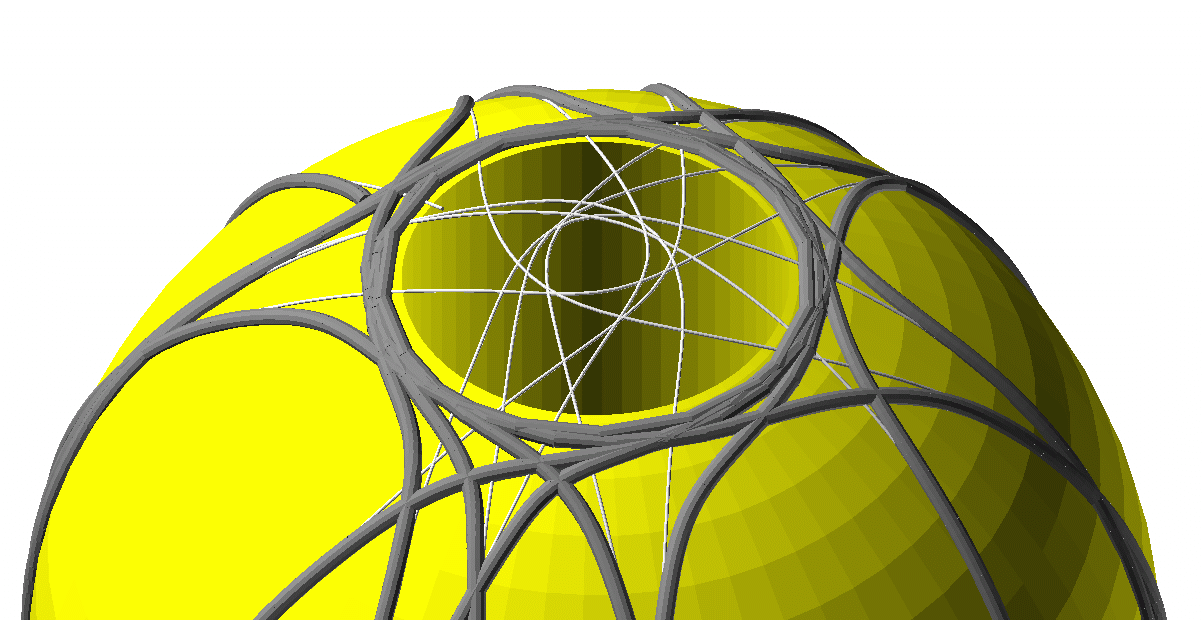} 
  \end{center}
\caption{Radial cloak
 on a spherical surface, 
 again based on a logarithmic function; 
 however unlike fig. \ref{fig-spherecloak}
 the original spherical surface has a varying 
 instead of a constant background index; 
 this perturbs the ray paths.
This figure shows only a long segment taken from a single ray path, 
 which returns to the vicinity of the north pole to 
 be cloaked (from) over and over again.
The thin lines show the path the ray would have taken
 if the cloak were not present.
}
\label{fig-sphexicloak}
\end{figure}

For example, 
 with a standard linear transformation where
 $\indevice{\theta} = \alpha \indesign{\theta} + \theta_0$, 
 and with $T \equiv T(f(\indevice{\theta}))$ and $P \equiv P(\indevice{\phi})$,
 where $\alpha=1-\theta_0/\pi$, 
 then the device metric must be
~
\begin{align}
  d\indevice{S}^2 
&=
  \left[
    \frac{T}{\alpha}
  \right]^2
  ~
  R^2 d{\indevice{\theta}}^2 
 +
  \left[
    P
    \frac{\sin \left( \frac{\indevice{\theta}-\theta_0}{\alpha} \right)}
         {\sin (\indevice{\theta})}
  \right]^2
  ~
  R^2 
  \sin^2 (\indevice{\theta})
  d\indevice{\phi}^2
.
\end{align}
An alternative cloaking deformation 
 based on the logarithmic one used in subsection \ref{S-examples-ccylinder}
 could be $\indevice{\theta}=\theta_0 \exp(\indesign{\theta}/e\theta_0)$.
This morphism has taken a partial spherical manifold
 and morphed it into a (near) full sphere; 
 thus the morphism applies only over
 $\indevice{\theta} \in (\theta_0,e\theta_0]$
 and (but) mimics the range $\indesign{\theta} \in (0,e\theta_0]$.
On the rest of the sphere, 
 i.e. for angles $\theta \in (e\theta_0,\pi]$, 
 we have that $\indevice{\theta}=\indesign{\theta}$.
This gives a morphed (T-device) metric
~
\begin{align}
  d\indevice{S}^2 
&=
  \left[
    \frac{T e\theta_0}{\indevice{\theta}}
  \right]^2
  R^2 d{\indevice{\theta}}^2 
\nonumber
\\
&\quad
 +
  \left[
    P
    \frac{\sin \left( e \theta_0 \log\left\{\frac{\indevice{\theta}}{\theta_0}\right\} \right)}
         {\sin (\indevice{\theta})}
  \right]^2
  R^2 
  \sin^2 (\indevice{\theta})
  d\indevice{\phi}^2
,
\end{align}
where since the metric is diagonal,
 the ${\theta\theta}$ and ${\phi\phi}$ 
 components of the inverse-metric $\indevice{\gMetric}^{ij}$, 
 speed squared $\indevice{\CSQ}^{ij}$,
 or diffusion ${\Dffsn}^{ij}$ matrices, 
 are just given by the inverse of square bracket terms.
These inverses are in essence just rescaling factors 
 for the speeds $c_\theta$ and $c_\phi$
 on the sphere in the angular and azimuthal directions.
Thus the logarithmically cloaked sphere, 
 in the transformed region,
 needs to have its material parameters modified so that
~
\begin{align}
  \frac{\indevice{c}_\theta} 
       {\indesign{c}_\theta}
=
  \sqrt{
    \frac{\indevice{\CSQ}^{\theta\theta}}
         {\indesign{\CSQ}^{\theta\theta}}
  }
&=
  \left[
    \frac{\indevice{\theta}}{e\theta_0}
  \right]
,
\\
  \frac{\indevice{c}_\phi} 
       {\indesign{c}_\phi}
=
  \sqrt{
    \frac{\indevice{\CSQ}^{\phi\phi}}
         {\indesign{\CSQ}^{\phi\phi}}
  }
&=
  \left[
    \frac{\sin (\indevice{\theta})}
         {\sin \left( e \theta_0 \log\left\{\frac{\indevice{\theta}}{\theta_0}\right\} \right)}
  \right]
.
\end{align}
As before,
 we can choose implement this anisotropic speed profile
 for either EM or acoustics following a similar procedure 
 as for the ordinary cylindrical cloak;  
 we might equally as easily follow the rules to work out the 
 material parameters needed for a heat diffusion cloak.

A depiction of this cloak, 
 implemented on a featureless sphere where $T=P=1$, 
 is shown in fig. \ref{fig-spherecloak}, 
 and showing a variety of deformed -- cloaked --
 great circle geodesic trajectories.
For more complicated spheres, 
 such as ones with a pre-existing index profiles
 that vary over the surface, 
 the geodesics will no longer be great circles.
Cloaking on such a sphere is displayed on fig. \ref{fig-sphexicloak}.
In this example a single ray trajectory will now travel widely
 over the surface in a complicated manner, 
 and so returns again and again to the north pole region 
 to be cloaked and recloaked in different ways
 and from different directions.

%
\subsection{Topographic transformation}

Imagine we wish to control our waves or rays
 so that they appear to be 
 travelling along
 a designer bumpy three dimensional landscape, 
 even though they remain confined to a planar device space, 
 albeit a plane with appropriately modulated properties.
If the height of the virtual landscape is defined by the function 
 $\indesign{z}=h(\indesign{x},\indesign{y})$, 
 then the required 2D metric that mimics it is 
 based on 
 $(\Morph^{\pstar}_C)\indices{^z_x} = \partial h/ \partial \indevice{x}$
 and
 $(\Morph^{\pstar}_C)\indices{^z_y} = \partial h/ \partial \indevice{y}$, 
 being
~
\begin{align}
  d\indesign{S}^2
&=
  \left[
    1 + \left(\frac{\partial h}{\partial {\indevice{x}}}\right)^2
  \right]
  d{\indevice{x}}^2
 +
  \left[
    1 + \left(\frac{\partial h}{\partial {\indevice{y}}}\right)^2
  \right]
  d{\indevice{y}}^2
\nonumber
\\
&\qquad\qquad\qquad
 +
  2
  \left[
    \left( \frac{\partial h}{\partial {\indevice{x}}}\right)
    \left(\frac{\partial h}{\partial {\indevice{y}}}\right)
  \right]
  d\indevice{x} ~d\indevice{y}
.
\end{align}

If, 
 for example
 we wished to mimic a parabolic or hyperbolic landscape, 
 defined by the height function 
 $h_1(\indevice{x},\indevice{y})=\alpha \indevice{x}^2 + \beta \indevice{y}^2$, 
 then the required device metric is 
~
\begin{align}
  d\indesign{S}^2
&=
  \left[
    1 + 4 \alpha^2 \indevice{x}^2
  \right]
  d{\indevice{x}}^2
 +
  \left[
    1 + 4 \beta^2 \indevice{y}^2
  \right]
  d{\indevice{y}}^2
\nonumber
\\
&\qquad\qquad\qquad
 +
  2
  \left[
    4
    \alpha \beta \indevice{x} \indevice{y}
  \right]
  ~d\indevice{x} ~d\indevice{y}
.
\end{align}
The metric components ${\gMetric}_{xx}$, 
 ${\gMetric}_{xy}$, 
 and ${\gMetric}_{yy}$ can then be read off directly from this result.
The components of ${\CSQ}^{ij}$ or ${\Dffsn}^{ij}$ 
 can then be determined, 
 being given by the \emph{inverse} of the matrix
~
\begin{align}
  \begin{bmatrix}
    {\gMetric}_{xx} & {\gMetric}_{xy} \\
    {\gMetric}_{xy} & {\gMetric}_{yy} 
  \end{bmatrix}
=
  \begin{bmatrix}
    1 + 4 \alpha^2 \indevice{x}^2 & 4 \alpha \beta \indevice{x} \indevice{y} \\
    4 \alpha \beta \indevice{x} \indevice{y} & 1 + 4 \beta^2 \indevice{y}^2
  \end{bmatrix}
.
\end{align}

Alternatively, 
 if we wished to mimic a landscape
 with a deep well (or peak) as 
 defined by the height function 
 $h_2(\indevice{x},\indevice{y})
  =\gamma/( 1 + \alpha \indevice{x}^2 + \beta \indevice{y}^2)$, 
 then the required device metric is 
~
\begin{align}
  d\indesign{S}^2
&=
  \left[
    1
   +
    \left( \frac{4\alpha^2}{\gamma^2} \right)
    \indevice{x}^2 
    ~ 
    h_2(\indevice{x},\indevice{y})^4
  \right]
  d{\indevice{x}}^2
\nonumber
\\
&\quad\qquad
 +
  \left[
    1 
   + 
    \left( \frac{4\beta^2}{\gamma^2} \right)
    \indevice{y}^2 
    ~ 
    h_2(\indevice{x},\indevice{y})^4
  \right]
  d{\indevice{y}}^2
\nonumber
\\
&\qquad\qquad\qquad
 +
  2
  \left[
    \left( \frac{4\alpha\beta}{\gamma^2} \right)
    \indevice{x} \indevice{y} 
    ~
    h_2(\indevice{x},\indevice{y})^4
  \right]
  ~d\indevice{x} ~d\indevice{y}
\label{eqn-example-topogpeak}
.
\end{align}
Of course, 
 many other landscapes can be imagined, 
 for example those considered when making 
 surface wave cloaks  
 \cite{MitchellThomas-MQHH-2013prl,MitchellThomas-QMHH-2014ol,Horsley-HMQ-2014sr}
 or geodesic lenses
 \cite{Sarbort-T-2012jo,Kinsler-TTTK-2012ejp,Luneberg-MTO}.

\begin{figure}
  \begin{center}
    \includegraphics[width=\figwidthFull,angle=0]{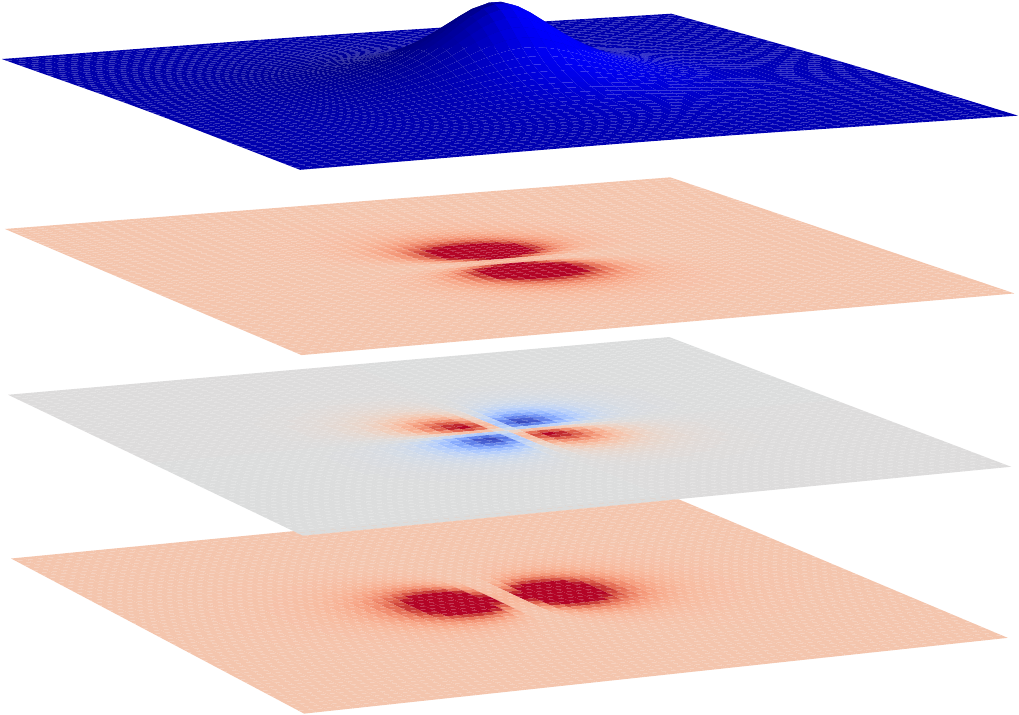}
  \end{center}
\caption{A 3D depiction of a topographic design manifold (top),
 given by \eqref{eqn-example-topogpeak}
 with $\alpha=\beta=20$ and $\gamma=1$.
The necessary metric properties 
 of the planar device (manifold), 
 i.e. the 
 $\indevice{x}\indevice{x}$,
 $\indevice{x}\indevice{y}$,
 and
 $\indevice{y}\indevice{y}$ components, 
 are given as color-coded surfaces below.
These metric components, 
 in a flat (planar) device, 
 mimic the distorted design manifold.
}
\label{fig-topography}
\end{figure}

%

This kind of landscape T-Design might lead us to consider the reverse case:
 can we, 
 by transformation, 
 modulate the properties of a bumpy
 but locally isotropic (device) sheet embedded in 3D -- 
 a pre-existing landscape of the type discussed above, 
 with height $z=h(x,y)$ 
 --
 so that it is designed to appear as if it were instead a flat sheet in 2D?
The answer is, 
 in general, 
 an emphatic no; 
 although {this} can be done in some specific cases: 
 i.e. the geodesic lenses mentioned above.

Imagine we have waves or rays travelling along 
 some kind of bumpy landscape that we wish to
 re-map to a flat space. 
Crucially,
 in some places, 
 for example, 
 the local curvature will cause some geodesics
 to converge at and through a focus.
Now no matter what diffeomorphism we apply,  
 we cannot remove that focus, 
 but only shift its position.
{Further, 
 since any collection of geodesics in a flat space
 can at most all share only a single focus,
 as soon as a landscape is such that if \emph{anywhere}
 a collection of geodesics share \emph{two} foci, 
 we cannot (in general) 
 diffeomorphically transform 
 from one to the other.}

%
\subsection{Focus transformation}

\begin{figure}
  \begin{center}
    \includegraphics[width=\figwidthFull,angle=0]{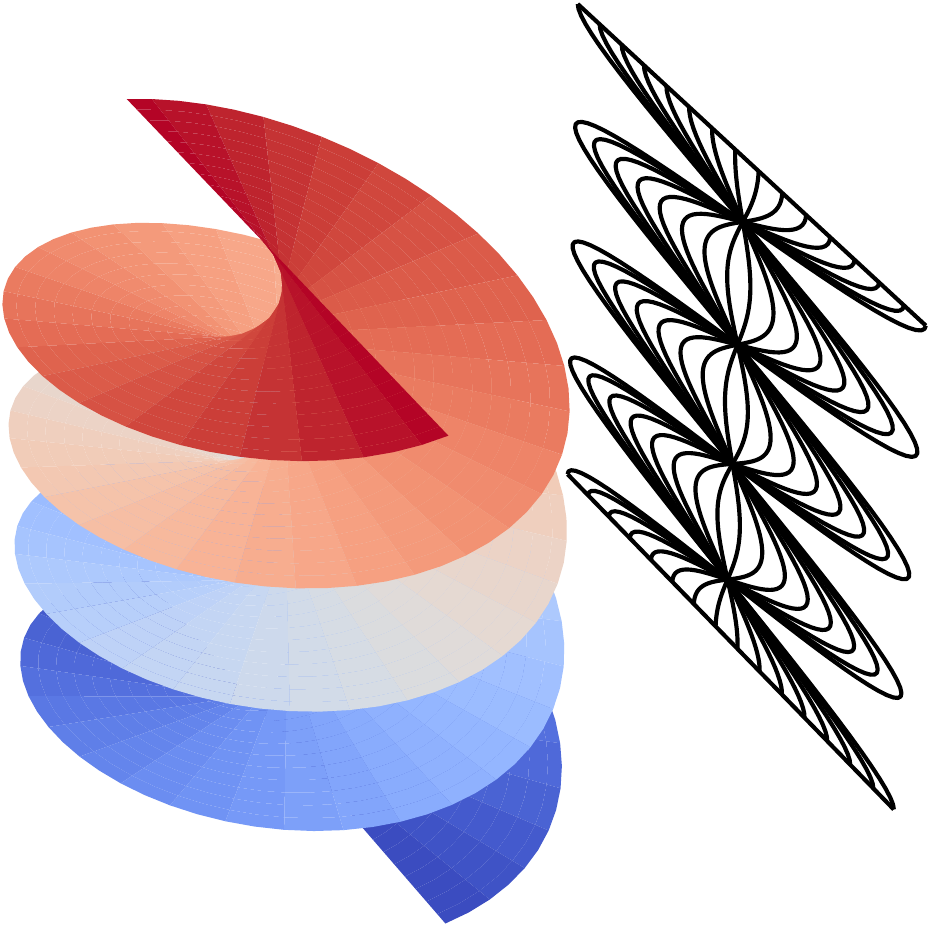}
  \end{center}
\caption{A 3D depiction of the helical design manifold
 used in the focus transformation, 
 with a $\indesign{z}$-dependent colour gradient to aid interpretation, 
 and the resulting ray trajectories
 in $\indevice{x}\indevice{z}$-plane of the device (manifold).
}
\label{fig-focustransform}
\end{figure}

We can imagine representing device that focuses in the 2D plane
 as a T-Design
 by embedding it in 3D and twisting the space along the focal axis, 
 as depicted in fig. \ref{fig-focustransform}.
With $z$ chosen as the focal axis, 
 points are twisted off the $x$-axis into the $xy$-plane, 
 using a rotation defined by 
~
\begin{align}
 \indesign{z} &= \indevice{z},  ~~\quad
 \indesign{x} = \indevice{x} \cos( \Phi \indevice{z} + \phi ),  ~~\quad
 \indesign{y} = \indevice{x} \sin( \Phi \indevice{z} + \phi)
.
\end{align}
The device space of $\indevice{z},\indevice{x}$
 will then mimic the behaviour 
 design space's twisted version embedded in 
 $\indesign{z},\indesign{x},\indesign{y}$.
This will involve periodic refocusings at $\Phi \indevice{z} + \phi = (2n+1) \pi /2$.
This T-Design specification means that the two spaces 
 differ in a nontrivial way (only) due to
 $(\Morph^{\pstar}_E)\indices{^x_x}$, 
 $(\Morph^{\pstar}_E)\indices{^x_z}$, 
 $(\Morph^{\pstar}_E)\indices{^y_x}$, 
 and
 $(\Morph^{\pstar}_E)\indices{^y_z}$.
These give
~
\begin{align}
  {d\indesign{z}} &= {d\indevice{z}},
\\
  {d\indesign{x}} &= -\Phi \indevice{x} \sin( \Phi \indevice{z} + \phi) {d\indevice{z}} + \cos( \Phi \indevice{z} + \phi) {d\indevice{x}},
\\
  {d\indesign{y}} &= ~~~\Phi \indevice{x} \cos( \Phi \indevice{z} + \phi) {d\indevice{z}} + \sin( \Phi \indevice{z} + \phi) {d\indevice{x}}
,
\end{align}
 and so 
~
\begin{align}
  {d\indesign{S}}^2
&=
  {d\indesign{x}}^2 + {d\indesign{y}}^2 + {d\indesign{z}}^2
\\
&=
  \left( 1 + \Phi^2 \indevice{x}^2 \right) {d\indevice{z}}^2
 +
  {d\indevice{x}}^2
,
\end{align}
 so we have 
 (almost) reinvented the parabolic index waveguide; 
 the difference being that only the $z$-directed index profile
 is modulated.
The resulting anisotropy means that -- 
 as is obvious from the transformation used -- 
 the structure preserves path lengths
 and will have no spatial dispersion in the ray limit.

The $x,z$ speed squared matrix 
 for the anisotropic material required for this device design 
 when applied to waves or rays 
 is then
~
\begin{align}
  {\CSQ}\indices{^{ij}} 
= 
  c^2
  \begin{bmatrix}
    1         & 0 \\
    0         & 1 + \Phi^2 \indevice{x}^2
  \end{bmatrix}^{-1}
.
\end{align}

Further, 
 we could note that this transformation 
 acts rather like a 2D projection of the helical transformation
 discussed by McCall et al.  \cite{McCall-KT-2016jo-helimed}.

Note that at a given focus point $\indevice{z}_i$, 
 the transformation projects multiple points 
 (actually the entire $\indesign{y}$-axis)
 down to a single point in the device, 
 namely $\indevice{z}=\indevice{z}_i$
 and $\indevice{x}=0$.
Indeed, 
 the device manifold consists of the $\indevice{x}\indevice{z}$-plane, 
 but with the set of all points consisting of the lines along $\indevice{z}= [(2n+1) \pi /2 - \phi]/\Phi$ except when $\indevice{x}=0$ removed.
Nevertheless, 
 rays passing through these foci are still distinguishable
 from each other by their direction.
Remarkably,
 we can also see that the device properties are insensitive to the
 chosen phase offset $\phi$.

%
\section{Conclusion}
\label{S-conclusions}

\TEXTNEW{Here we have shown
 the extent to which all the distinct types
 of transformation design might be repackaged into a general formalism.
Although this process has necessarily involved approximations,
 we have shown that it is possible
 to make a clear distinction between
 the mathematical design step
 and the subsequent choice of which physical model 
 is used to implement it.}{(6)}

Indeed,
 from the perspective given here, 
 there is absolutely nothing ``magic'' about transformation optics,  
 acoustics,
 or any of the other transformation domains -- 
 as long as we are prepared to tolerate approximations.
If we have a pre-specified metric,
 then we can map this directly to a speed-squared matrix 
 and use our knowledge of materials 
 (or of metamaterials)
 to work out an implementation.
Alternatively,
 and this is the usual case, 
 if we have a useful scheme for reconfiguring the flow or location
 of the light or acoustic waves (or rays), 
 then we can simply transform our design (``reference'') metric --
 usually the vacuum, but this is not a requirement -- 
 directly into the necessary device metric.
This process involves the calculation 
 of only a couple of matrix multiplications
 at each point in the transformed domain.
Then, 
 as before, 
 the demands of any specific implementation are straightforward to identify.

Notably, 
 we can see that if we eschew issues of impedance handling
 and changes in volume measure 
 the wave and ray transformation procedures 
 to be used are the same.
This is not to deny the importance of impedance, 
 merely to note the similarities between transformations 
 of waves and rays.
Further, 
 the ``obvious'' process of matched modulation
 of both constitutive parameters --
 e.g. $\epsilon$ and $\mu$ in optics, 
 $\kappa$ and $\rho$ in acoustics 
  -- 
 gives the natural choice of impedance mismatch, 
 even if the design intent is only for a ray T-device.

As it stands here, 
 we only consider purely spatial transformations.
However, 
 it has already been shown 
 \cite{McCall-FKB-2011jo,Kinsler-M-2014adp-scast,Kinsler-M-2014pra}
 that an extension to spacetime transformations 
 can be done in a relatively straightforward way, 
 at least in the 1+1D case.
In contrast, 
 spacetime transforms of dispersive \cite{Gratus-KMT-2016njp-stdisp}
  and diffusive systems
 are more problematic, 
 and is an area we are actively investigating.

%
\section*{Acknowledgments}
We acknowledge valuable discussions with 
 Robert Thompson
 and David Topf;
 as well as funding from
 EPSRC grant number EP/K003305/1.
{PK would also like to acknowledge recent support from 
 the EPSRC/Alpha-X grant EP/J018171/1
 and STFC grant G008248/1.}


\section*{Appendix: Transforming Metrics}

\def\comet{{g}}
\def\covec{{u}}
\def\morph{\varphi}
\def\point{{\cal P}}
\def\minkmetric{\eta_{\bar\alpha\bar\beta}}
\def\fieldM{{\cal \chi}({\cal M})}
\def\cofieldM{{\cal\chi}^*(\cal M)}
\def\coder{{\boldsymbol \nabla}}
\def\newcoder{{\tilde{\boldsymbol \nabla}}}
\def\newGamma{{\tilde {\Gamma}}}
\def\intcoder{{\bar{\boldsymbol \nabla}}}
\def\intGamma{{\bar \Gamma}}
\def\newg{{\tilde g}}
\def\Xpara{X_\parallel}
\def\Xperp{X_\perp}
\def\newX{{\tilde X}}
\def\newY{{\tilde Y}}
\def\newN{{\tilde N}}
\def\newx{{\tilde x}}
\def\newr{{\tilde r}}
\def\newtheta{{\tilde \theta}}
\def\newx{{\tilde x}}
\def\newy{{\tilde y}}
\def\blkmod{\kappa}
\def\phase{{\cal S}}
\def\curve{{\cal C}}
\def\mani{{\cal M}}
\def\nani{{\cal N}}
\def\real{{\mathbb R}}
\def\xtilde{{\tilde x}}
\def\ytilde{{\tilde y}}
\def\ztilde{{\tilde z}}

\setlength{\parindent}{0pt}
\setlength{\parskip}{10pt}

Take a vector space $V$ with a basis $\{{\bf e}_\alpha\}$. A basis of the dual space $V^*$ is a set of co-vectors $\{ {\boldsymbol \omega}^\beta \}$ satisfying ${\boldsymbol \omega}^\beta ({\bf e}_\alpha)=\delta^\beta_{~\alpha}$.

Take an $n$-dimensional manifold $\cal M$ with a coordinate chart $\phi: {\cal U}\to {\mathbb R}^n$, i.e. $\phi (p)=\{x^\alpha (p)\}$.

A {\em coordinate basis} $\{\partial_\alpha \}$ at $p\in {\cal M}$ is associated with a set of coordinate functions $x^\alpha : {\cal U} \to {\mathbb R}$.

Take a function $f: {\cal M} \to {\mathbb R}$. Tangent vectors in $T_p{\cal M}$, the tangent (vector) space at $p$,  act on such functions, to produce the number $v(f)$. With respect to the coordinate basis, $v(f)=v^\alpha \partial_\alpha f$.

If $T^*_p{\cal M}$ is the co-tangent space at $p$, then the {\em 1-form} $df\in T^*_p{\cal M}$ is defined via $df(v)=v(f)$. If we define $\{dx^\beta\}$ to be the basis dual to $\{ \partial_\alpha \}$ (i.e. $dx^\beta (\partial_\alpha)=\delta^\beta_{~\alpha}$), then $df=\partial_\alpha f dx^\alpha$.

A {\em diffeomorphism} $\varphi: {\cal M} \to {\cal N}$ induces the {\em pull-back} of functions from $\cal N$ to $\cal M$ according to $\morph^*f (p) = f\circ \morph (p)$.

The diffeomorphism also induces the {\em push-forward} of vectors $v\in T_p{\cal M}$ according to $\morph_*v(f)=v(\morph^*f)$. In components $(\morph_* v)^\mu \partial_\mu = v^\alpha(\morph_*)^\mu_{~\alpha}\partial_\mu$, where $(\morph_*)^\mu_{~\alpha}=\partial y^\mu/\partial x^\alpha$, and $y^\mu$ is a coordinate function on $\cal N$ associated with $\morph (p)$. It also induces the pull-back of 1-forms $df\in T^*_{\morph(p)}{\cal N}$ according to $(\morph^*df) (v)=df(\morph_*v)$.\footnote{Incidentally $(\morph^*df)(v)=df(\morph_*v)=(\morph_*v)(f)=v(\morph^*f)=d(\morph^*f)(v)$, so that $[d,\morph^*]=0$.} In components $(\morph^*df)_\alpha dx^\alpha = (df)_\mu (\morph^*)^\mu_{~\alpha}dx^\alpha$, where $(\morph^*)^\mu_{~\alpha}=\partial y^\mu/\partial x^\alpha$. Note that $(\morph_*)^\mu_{~\alpha}=(\morph^*)^\mu_{~\alpha}$. The positioning of the $*$ indicates whether we are pushing forward a vector or pulling back a 1-form.

A metric on ${\cal N}$ is a symmetric bilinear function of vectors, i.e. $g:T_p{\cal N}\times T_p{\cal N} \rightarrow {\mathbb R}$. If $\{dy^\mu\}$ is a coordinate co-basis of forms spanning $T^*_{\morph (p)}{\cal N}$, then $g=g_{\mu\nu}dy^\mu \otimes dy^\nu$, where $g_{\mu\nu}=g_{\nu\mu}$. This metric can be pulled-back to $\cal M$ via $\morph^*g (u,v)=g(\morph_*u,\morph_*v)$, where $u,v \in T_p{\cal M}$. Denoting this induced metric as ${\tilde g}=\morph^*g$ we have, with respect to coordinate bases at $p\in \cal M$ and $\morph (p)\in \cal N$, ${\tilde g}_{\alpha\beta}=(\morph^*)^\mu_{~\alpha}(\morph^*)^\nu_{~\beta}g_{\mu\nu}$.

A {\em co-metric} on $\cal M$ is a symmetric bilinear function of co-vectors, i.e. $\comet:T^*_p{\cal M}\times T^*_p{\cal M} \rightarrow {\mathbb R}$. 

If $\{\partial_\alpha\}$ is a coordinate basis of vectors spanning $T_p{\cal M}$, then $\comet = g^{\alpha\beta}\partial_\alpha \otimes\partial_\beta$, where $g^{\alpha\beta}=g^{\beta\alpha}$. This co-metric can be pushed-forward to $\cal N$ via $\morph_*\comet (\lambda,\rho)=\comet(\morph^*\lambda,\morph^*\rho)$, where $\lambda,\rho\in T^*_{\morph(p)}{\cal N}$. Denoting this induced co-metric as ${\tilde \comet}=\morph_*\comet$ we have, with respect to coordinate bases in $\cal M$ and $\cal N$, ${\tilde g}^{\mu\nu}=(\morph_*)^\mu_{~\alpha}(\morph_*)^\nu_{~\beta}g^{\alpha\beta}$.

Now let's work in a single manifold $\cal M$ equipped with a metric $g$. A vector $u\in T_p{\cal M}$ can be assigned a `squared length' $u^2=g(u,u)$. In $T^*_p{\cal M}$ choose a co-vector $\covec$ such that $\covec (u)=g(u,u)$. A natural choice of co-metric is then one which sets $\comet(\covec,\covec)=g(u,u)=\covec (u)$. In that case it is easily shown that $u=\comet(~,\covec)$, $\covec=g(u,~)$ and $\comet^{\alpha\gamma}g_{\gamma\beta}=\delta^\alpha_{~\beta}$, where $\{e_\beta\}$ and $\{\omega^\alpha\}$ are respectively a basis of $T_p{\cal M}$ and a co-basis of $T^*_p{\cal M}$, i.e. $\omega^\alpha(e_\beta)=\delta^\alpha_{~\beta}$. It then makes sense to refer to $\comet$ as the inverse of $g$.

Summary: you can pull-back a metric from $\cal N$ to $\cal M$, and you can push-forward a co-metric from $\cal M$ to $\cal N$. In a space equipped with a metric $g$ there is a natural co-metric $\comet$ which is the inverse of $g$ (i.e. $g$ takes $u$ to $\covec$, and $\comet$ takes $\covec$ back to $u$).

\end{document}